\documentclass[usenatbib]{mn2e}
\usepackage{psfig}

\newcommand{\etal}{et~al.}  
\newcommand{\ionhy}{H{\sc ii}}
\newcommand{\degrees}{$^\circ$} 
\newcommand{\kms}{$\mbox{km~s}^{-1}$}
\newcommand{\water}{$\mbox{H}_{2}\mbox{O}$}

\newcommand{\specdfig}[2]        % Double FIGures (put two figures side by
{
  \begin{center}
    \begin{minipage}[t]{0.45\textwidth}
        \psfig{file=./#1.eps,height=0.75\textwidth,angle=270}
    \end{minipage}
    \hfill
    \begin{minipage}[t]{0.45\textwidth}
        \psfig{file=./#2.eps,height=0.75\textwidth,angle=270}
    \end{minipage}
  \end{center}
}

\newcommand{\specsfig}[1]        % Single FIGure (put one figure of the same
{
  \begin{center}
    \begin{minipage}[t]{0.45\textwidth}
        \psfig{file=./#1.eps,height=0.75\textwidth,angle=270}
    \end{minipage}
  \end{center}
}

\newcommand{\speclfig}[2]        % Single FIGure (put one figure of the same
{
  #1
  \begin{center}
    \begin{minipage}[t]{0.45\textwidth}
      \psfig{file=./#2.eps,height=0.75\textwidth,angle=270}
    \end{minipage}
  \end{center}
}

\begin{document}

\title[Class~I \& Class~II Methanol Masers] 
{The Relationship between Class~I and Class~II Methanol Masers}

\author[Ellingsen]{S.P. Ellingsen$^1$\\
$^1$ School of Mathematics and Physics, University of Tasmania, 
     Private Bag 21, Hobart, Tasmania 7001, Australia;\\  
     Simon.Ellingsen@utas.edu.au}

\maketitle

\begin{abstract}
  
  The Australia Telescope National Facility Mopra millimetre telescope
  has been used to search for 95.1-GHz class~I methanol masers towards
  sixty-two 6.6-GHz class~II methanol masers.  A total of twenty-six
  95.1-GHz masers were detected, eighteen of these being new
  discoveries.  Combining the results of this search with observations
  reported in the literature, a near complete sample of sixty-six
  6.6-GHz class II methanol masers has been searched in the 95.1-GHz
  transition, with detections towards 38 per cent (twenty-five
  detections ; not all of the sources studied in this paper qualify
  for the complete sample, and some of the sources in the sample were
  not observed in the present observations).
  
  There is no evidence of an anti-correlation between either the
  velocity range, or peak flux density of the class~I and II
  transitions, contrary to suggestions from previous studies.  The
  majority of class~I methanol maser sources have a velocity range
  that partially overlaps with the class~II maser transitions.  The
  presence of a class~I methanol maser associated with a class~II
  maser source is not correlated with the presence (or absence) of
  main-line OH or water masers.  Investigations of the properties of
  the infrared emission associated with the maser sources shows no
  significant difference between those class~II methanol masers with
  an associated class~I maser and those without.  This may be
  consistent with the hypothesis that the objects responsible for
  driving class~I methanol masers are generally not those that produce
  main-line OH, water or class~II methanol masers.
  
\end{abstract}


\section{Introduction}

Intense activity in the field of methanol maser research commenced in
the mid 1980s with the discovery of numerous new transitions
\citep[e.g.][and others]{MOK85,WWSJ84,WWMH85}.  As the number of
methanol maser transitions increased it became clear that they could
be classified empirically into two groups on the basis of the sources
towards which they were detected \citep{BMMW87,M91a}.  Class~II
methanol masers are found close to sites of high-mass star formation
and are often associated with \ionhy\ regions, infrared sources and OH
masers.  The strongest and most widespread class~II transitions are
the $5_1 - 6_0\mbox{A}^{+}$ (6.6~GHz) and $2_0 - 3_{-1}\mbox{E}$
(12.1~GHz), which have been detected towards more than 500 sites
within the Galaxy \citep*[][and references therein]{PMB04}.  The
archetypal class~II methanol maser source is W3(OH).  Class~I methanol
masers are also found towards star formation regions, but offset by
small, though significant amounts from \ionhy\ regions, infrared
sources, OH and water masers.  The strongest class~I transitions are
the $7_0 - 6_1\mbox{A}^{+}$ (44.0~GHz) and $4_{-1} - 3_0\mbox{E}$
(36.1~GHz), and the archetypal source is Orion~KL.  The class~II $2_0
- 3_{-1}\mbox{E}$ transition is often observed in absorption towards
class~I sources \citep{PW92}.  Theoretical models of methanol masers
provide some explanation for this empirical division predicting
inversion in the class~I transitions when collisional processes
dominate and inversion in the class~II transitions when radiative
processes dominate \citep*{CJGB92,SD94,SCG97}.

The classification suggested by \citet{M91a} was consistent with the
observational and theoretical understanding at the time. However, more
recent studies suggest that things may be more complicated.  A number
of searches for class~I methanol masers have been made towards
traditional class~II sites with relatively high detection rates
\citep{SKVO94,VES+00}.  \citet{SKVO94} suggested an anti-correlation
between the intensities and velocity ranges of the class~I and II
methanol masers within the same region.  This requires that there is
some sort of relationship between the emission from the two
transitions since, if they were independent neither correlations, nor
anti-correlations should exist.  Perhaps more importantly, modelling
predicts that in some conditions the 6.6-GHz class~II transition can
be weakly inverted, simultaneously with strong inversion in the 25-GHz
class~I transitions.  This situation may have recently been detected
towards the archetypal class~I maser source Orion~KL \citep{VSEO04}.
\citet*{KHA04} found coincidence within 0.5 arcsec of class~I and II
methanol masers in one source, although there is a significant
velocity difference between the two transitions suggesting that it may
be a chance alignment along our line of sight.

The majority of studies of methanol masers have focused upon the
class~II transitions.  There are two main reasons for this ; the first
is that the strongest class~II transitions are in the centimetre
regime and hence easier to observe ; the second is that their
association with other maser species, {\em IRAS} sources etc makes
targeted searches relatively productive
\citep[e.g.][]{M91b,CVEWN95,WHRB97}.  Although untargeted searches
have also been very successful at detecting class~II methanol masers
\citep{C96,EVM+96,SKHKP02}.  In contrast the strongest class~I
methanol maser transitions are at millimetre wavelengths and to date
there is no other type of astrophysical object with which they are
known to be closely associated.  \citet{M91a} suggested that class~I
masers are offset from class~II sites by up to 1~pc.  However,
there are relatively few sources where both class~I and II transitions
have been imaged at high resolution.  The majority of the well studied
class~I methanol maser sources are ``traditional'' class~I sources
with no (or weak) class~II emission, Orion~KL \citep{PW88,JGS+92},
DR21 \citep{BM88,PM90} and Sgr~B2 \citep{MM97}.  There are no strong
class~II methanol masers towards Orion~KL, although weak masers, and
possibly quasi-thermal emission has been detected \cite{VSEO04}.
There is class~II methanol maser emission towards DR21(OH), this
likely to be from the same region as the OH maser emission (which is
offset from the class~I masers), but this has not been confirmed.
Sgr~B2 is a very complex region, with multiple site of both class~I
and II methanol masers.  The class~II masers are close to the compact
\ionhy\ regions \cite{HW95}, while the majority of class~I masers lie
in an arc, possibly tracing the interface between two molecular clouds
\cite{MM97}.

For a source at a distance of 3~kpc (typical for a high-mass star
forming region) a linear distance of 1~pc corresponds to an angular
separation of slightly more than 1 arcminute.  This is comparable to,
or larger than typical telescope beam sizes at millimetre wavelengths
and so searches targeted at class~II maser sites
\citep[e.g.][]{SKVO94,VES+00} would be expected to detect few class~I
sources.  This is not the case though, which suggests that in many
cases the offset between class~I and II maser sites is significantly
less than 1~pc.  This has been confirmed by recent VLA observations of
44~GHz class~I methanol masers \citep{KHA04} which found a median
separation of 0.2~pc from \ionhy\ regions.  At the present time,
class~II methanol maser sites appear to be the best targets for
class~I maser searches.  However, the nature of the relationship
between the two classes remains unclear.  To try and elucidate the
nature of the relationship I have undertaken a search for class~I
methanol masers towards a statistically complete sample of class~II
methanol masers selected from the Mt Pleasant survey
\citep{EVM+96,E96}.

\section{Observations and Data reduction} \label{sec:obs}

The observations were made between 1998 July 12--17 using the
Australia Telescope National Facility (ATNF) 22m millimetre antenna at
Mopra.  At the time of the observations only the inner 15~m of the
Mopra antenna was illuminated and the antenna had a sensitivity of
approximately 40 Jy\ K$^{-1}$.  The assumed rest frequency of the $8_0
- 7_1\mbox{A}^{+}$ was 95.169~489~GHz \citep{DHAH89} and at this
frequency the half-power beam width of the Mopra antenna was
52~arcsec.  Recent measurement of the rest frequencies of various
methanol maser transitions by \citet*{MMM04} gives a value of
95.169~463~GHz (with an uncertainty of 10~kHz).  This is 14~kHz lower
than the value used, which means that the velocities listed are
0.044~\kms\ higher than had the more recent rest frequency value been
used.  The data were collected using a 2-bit digital autocorrelation
spectrometer configured with 1024 channels spanning a 32-MHz
bandwidth.  For an observing frequency of 95.1~GHz this configuration
yields a natural weighting velocity resolution of 0.12 \kms, or
0.2~\kms after Hanning smoothing.  The SIS mixers on both of the
available linear polarization channels were tuned to the 95.1~GHz
methanol transition.  The antenna pointing was checked and corrected
prior to each observing session by making observations of 86-GHz SiO
masers; the nominal pointing accuracy is of the order of 10 arcsec
RMS.  For one of the channels the tuning was performed only once at
the beginning of the observing period.  However, the other channel had
to be re-tuned each day after the pointing checks.  The sensitivity of
this channel varied day to day depending upon how well the receiver
tuned, in general it had a system temperature approximately 50K higher
than the other channel, though on occasions it was slightly lower.
The system temperature was determine by inserting an ambient
temperature load which was assumed to have a temperature of 295K.  The
measured system temperature varied between 200 and 340K depending upon
the weather conditions and elevation.  This method of calibration also
corrects for atmospheric absorption \citep{KU81} and taking into
account pointing inaccuracies the absolute accuracy of the flux
density scale of the observations is conservatively estimated to be 20
per cent.  The sources were observed in a position switching mode with
300 seconds spent at the on-source position and 300 seconds at a
reference position offset in declination 30~arcminutes to the south.
This procedure was repeated 3 times to yield a total on-source
integration time of 15 minutes for most sources, which typically
yielded an RMS of 1.3~Jy after averaging the two polarizations and
Hanning smoothing.  \citet*{WTW04} recently observed linear
polarization in excess of 10\% in a number of class~I methanol maser
transitions, including the 95.1-GHz.  The spectra presented here are
an average of two orthogonal linear polarizations and hence the
relative flux density of the features will not be effected by any
linear polarization.  The current observations were made in such a way
that it is not possible to determine the polarization characteristics
of the masers from them.  The data were processed using the {\sc spc}
reduction package.  Quotient spectra were formed for each on/off pair
of observations which were then averaged together, a polynomial
baseline fitted and subtracted and the velocity and amplitude scale
calibrated.  Velocity calibration presented some challenges, requiring
correction for incorrectly recorded frequency synthesiser chain
information in the RPFITS header \cite{LPPK04} and editing of the
telescope identity and band-inversion in the headers.

The sample of 6.6-GHz methanol maser sites searched for 95.1-GHz maser
emission was drawn from \citet{EVM+96} and \citet{E96} and included
all sources detected in the regions with Galactic longitude $l =
25^{\circ}$ -- $30^{\circ}, 282^{\circ}$ -- $286^{\circ}, 291^{\circ}$
-- $296^{\circ}, 325^{\circ}$ -- $335^{\circ}$.  For the first and
last of these longitude ranges the Galactic latitude range of the Mt
Pleasant survey was $b = -0\fdg53$ -- $+0\fdg53$, while for the other
regions it was $b = -1\fdg03$ -- $+0\fdg03$.  The total area covered
is approximately 24.4 square degrees.  The exact number of class~II
methanol masers sites within these regions depends on the spatial
resolution of the observations.  A number of the sources have two or
more sites of emission separated by a few arcseconds.  Such sources
cannot be resolved with the Mopra telescope and so in general we have
made observations towards individual sites only where they are
separated by more than 20 arcsec.  Observations of the 95.1-GHz
transition were made of a small number of sources not detected as part
of the original Mt Pleasant survey.  In particular,
$326.475\!+0\!.703$ which was listed as $326.40\!+\!0.51$ in the
original survey \citep{EVM+96}, the correct position being revealed by
Australia Telescope Compact Array (ATCA) observations.
$27.286\!+\!0.151$ is outside the velocity range of the Mt Pleasant
survey, but was discovered during ATCA observations of the nearby
source $27.223\!+\!0.137$.  In addition, the untargeted search of
\citet{C96} detected a number of 6.6-GHz methanol masers in the region
$l = 330^{\circ} - 335^{\circ}$ which were not detected in the Mt
Pleasant survey either due to variability or being outside the
searched velocity range ($331.120\!-\!0.118$, $333.029\!-\!0.063$,
$333.646\!+\!0.058$ \& $334.935\!-\!0.098$) and these were also
searched for 95.1-GHz methanol masers.  An independent untargeted
survey of the $l=20^{\circ} - 40^{\circ}$ region has recently been
undertaken with the Torun telescope \citep{SKHKP02}, which also
discovered a number of sources in the $l=25^{\circ} - 30^{\circ}$
region not detected at Mt Pleasant.  However these were not searched
as the information was not available at the time of the observations.
For the majority of sources (80 of 84) the positions search for
95.1-GHz class~I methanol masers are the sites of class~II masers
determined from ATCA observations which have a positional accuracy of
approximately 0.5 arcsec (Ellingsen, unpublished observations).  The
only exceptions were the sources $285.32\!-\!0.03$, $293.84\!-\!0.78$,
$293.95\!-\!0.91$ and $25.53\!+\!0.38$, for which the positions
determined in the Mt Pleasant survey (positional accuracy RMS 0.6
arcmin) were used \citep{EVM+96}.  For those sources where ATCA
position are available we have used three significant figures for the
Galactic coordinate names.  These names differ slightly from those
used previously in the literature, but the greater accuracy of the
ATCA positions warrants it, and the correspondence is in most cases
reasonably obvious.

A small number of the detected 95.1-GHz methanol masers were either
not observed, or have poor spectra from the original 6.6-GHz Mt
Pleasant survey.  To enable proper comparison of the class~I and
class~II emission in all class~I detections, new 6.6-GHz observations
were made of the sources 326.474$\!+\!$0.703, 333.029$\!-\!$0.063 and
27.286$\!+\!$0.151.  These observations were made with the Mt Pleasant
26-m telescope on 2004 November 3 \& 8.  The assumed rest frequency of
the $5_1 - 6_0\mbox{A}^{+}$ transition was 6.668518~GHz, and at this
frequency the half-power beam width of the Mt Pleasant antenna is
7~arcmin.  The observations were made with a cryogenically cooled
receiver with dual circular polarizations and a system equivalent flux
density of approximately 1200~Jy.  The data were collected using a
2-bit autocorrelation spectrometer configured with 4096 channels
spanning a 4-MHz bandwidth for each polarizations.  For an observing
frequency of 6.6-GHz this configuration yields a velocity resolution
after Hanning smoothing of 0.09~\kms.  Each source was observed in
position switching mode with 10 minutes spent at the on-source
position and a further 10 minutes at a reference position offset by
-1~degree in declination.  To measure any pointing errors a 5-point
grid, centred on the nominal position was observed for each source.
In each case the measured pointing offsets were small, implying
scaling corrections to the flux density scale of the source spectra of
less than 5~per cent.  As this is less than the accuracy of the flux
density calibration, no scaling was applied.

\section{Results}

A total of fifty-nine 6.6-GHz class~II methanol maser sites were
searched for 95.1~GHz class~I masers.  Emission was detected towards
twenty-six sites and these are listed, along with the parameters of
Gaussian fits to the emission in Table~\ref{tab:det95}.  Spectra of
each of the detected 95.1-GHz masers are shown in
Figure~\ref{fig:meth95}.  Eighteen of the 95.1-GHz emission sites
detected are new discoveries, with the remainder previously observed
by \citet{VES+00}.  The sources for which no emission was detected are
listed in Table~\ref{tab:nondet95}, along with the velocity range
searched and the 3$\sigma$ detection limit for the observation.

\begin{table*}
  \caption{Class~II methanol maser sites detected in the 95.1-GHz class I 
    transition.  The Hanning smoothed spectra have been fitted with one or more
    Gaussian profiles and the fitted parameters are listed, the number in
    brackets is the uncertainty reported by the fitting procedure.  Notes : 
    $^{a}$ sources not within the statistically complete sample of 6.6-GHz 
    methanol masers.}  
  \begin{tabular}{lllrrr} \hline
    {\bf Name} & {\bf Right Ascension} & {\bf Declination} & 
    {\bf Peak Flux} & {\bf Velocity} & {\bf Full Width}  \\
               & {\bf (J2000)}         & {\bf (J2000)}     & 
    {\bf Density}   & {\bf (\kms\/)} & {\bf Half Maximum} \\ 
               & {\bf (h~m~s)}      & {\bf (\degrees\/ \arcmin\/ \arcsec\/)} &
    {\bf (Jy)}      &                & {\bf (\kms\/)} \\ [2mm] \hline
    $328.237\!-\!0.548^{a}$ & 15:57:58.381 & -53:59:23.14 &  5.2(1.3) & 
      -41.8(0.1) & 1.6(0.3) \\ 
     & & & 4.4(0.8) & -41.3(0.05) & 0.4(0.1) \\
     & & & 3.8(0.5) & -44.0(0.2) & 2.2(0.3)  \\
     & & & 2.3(0.4) & -39.7(0.9) & 3.9(1.3)  \\
    $328.809\!+\!0.633^{a}$ & 15:55:48.608 & -52:43:06.20 &  18.3(1.6) & 
      -40.9(0.1) & 1.5(0.1) \\ 
     & & & 17.3(1.4) & -41.4(0.05) & 5.2(0.2)  \\
     & & & 14.1(1.2) & -40.5(0.05) & 0.4(0.05) \\
     & & & 13.9(1.1) & -42.6(0.05) & 1.3(0.1)  \\
    $329.031\!-\!0.198$     & 16:00:30.326 & -53:12:27.35 & 13.4(1.0) &
      -47.0(0.05) & 0.7(0.1) \\
     & & & 12.0(1.9) & -43.9(0.1) & 1.3(0.2)  \\
     & & & 8.0(0.5) & -46.0(0.2)  & 2.4(0.3)  \\
     & & & 6.7(2.4) & -43.5(0.1)  & 1.2(0.3)  \\
     & & & 5.7(2.0) & -43.7(0.05) & 0.4(0.2)  \\
     & & & 4.4(0.6) & -41.6(0.6)  & 4.4(0.9)  \\ 
    $331.132\!-\!0.244$     & 16:10:59.743 & -51:50:22.70 & 31.6(0.6)  &
      -91.1(0.05) & 0.8(0.05)  \\
     & & & 8.8(0.7) & -88.5(0.05) & 0.5(0.1)  \\
     & & & 8.4(0.2) & -86.7(0.1)  & 6.9(0.2)  \\
     & & & 5.6(0.6) & -84.6(0.05) & 0.8(0.1)  \\
    $331.342\!-\!0.346$     & 16:12:26.456 & -51:46:16.86 & 12.9(0.7) & 
      -65.8(0.05) & 0.6(0.05)  \\
%*** Use information from V00 observation?
    $332.295\!-\!0.094$ & 16:15:45.381 & -50:55:53.85 & 14.5(5.4) & 
      -50.8(0.05) & 0.5(0.1) \\ 
    $332.942\!-\!0.686^{a}$ & 16:21:19.018 & -50:54:10.41 &  4.6(0.4) &
      -48.6(0.1) & 3.6(0.3)  \\
      & & & 3.8(0.7) & -49.0(0.1) & 0.7(0.2) \\
    $332.963\!-\!0.679^{a}$ & 16:21:22.926 & -50:52:58.71 &  9.1(0.7) & 
      -46.8(0.05) & 0.6(0.1) \\
     & & & 8.5(1.0) & -47.4(0.05) & 0.5(0.1) \\
     & & & 3.0(0.3) & -49.0(0.2)  & 4.4(0.5) \\
    $333.029\!-\!0.063^{a}$ & 16:18:56.735 & -50:23:54.17 &  3.7(1.0) & 
     -41.0(0.1) & 0.4(0.1) \\
     & & & 2.8(0.5) & -40.1(0.1) & 1.0(0.3) \\
    $333.121\!-\!0.434$     & 16:20:59.704 & -50:35:52.32 &  6.1(0.2) & 
     -50.9(0.1) & 7.1(0.2) \\
     & & & 6.0(0.5) & -52.9(0.05) & 0.9(0.1) \\
     & & & 5.3(0.7) & -50.7(0.05) & 0.4(0.1) \\
    $333.128\!-\!0.440$     & 16:21:03.300 & -50:35:49.75 &  23.3(0.7) &
     -47.6(0.05) & 2.7(0.1) \\
     & & & 21.9(1.2) & -48.6(0.05) & 1.2(0.1) \\
     & & & 9.2(0.8) & -49.9(0.1) & 2.0(0.2) \\
     & & & 7.4(0.4) & -50.6(0.2) & 7.1(0.4) \\
    $333.130\!-\!0.560^{a}$ & 16:21:35.742 & -50:40:51.29 &  7.6(0.6) &
     -59.1(0.05) & 1.4(0.1) \\
     & & & 6.4(0.6) & -54.4(0.05) & 0.5(0.1) \\
     & & & 4.6(0.2) & -56.5(0.2) & 3.8(0.4)  \\
    $333.163\!-\!0.101^{a}$ & 16:19:42.670 & -50:19:53.20 &  3.7(0.5) &
     -91.3(0.05) & 0.6(0.1) \\ 
     & & & 3.5(0.7) & -109.8(0.05) & 0.5(0.1) \\
    $333.184\!-\!0.091$     & 16:19:45.620 & -50:18:35.00 &  4.4(1.8) &
     -87.0(0.1) & 0.2(0.2) \\
     & & & 4.3(0.8) & -86.5(0.1) & 1.4(0.3)  \\
    $333.234\!-\!0.062$     & 16:19:51.250 & -50:15:14.10 &  131.9(0.9) &
     -87.3(0.05) & 1.1(0.05) \\
     & & & 17.7(0.8) & -87.6(0.05) & 4.0(0.1) \\
    $333.315\!+\!0.105$     & 16:19:29.016 & -50:04:41.45 &  2.5(0.3) &
     -46.5(0.3) & 3.5(0.8) \\
     & & & 2.2(0.8) & -44.4(0.2) & 1.0(0.5) \\
    $333.466\!-\!0.164$     & 16:21:20.180 & -50:09:48.60 &  7.5(0.4) &
     -42.9(0.1) & 3.0(0.2) \\
     & & & 5.1(0.7) & -45.2(0.05) & 0.5(0.1) \\
     & & & 4.7(0.8) & -42.8(0.05) & 0.4(0.1) \\
    $333.562\!-\!0.025$     & 16:21:08.797 & -49:59:48.26 &  11.8(1.6) &
     -40.0(0.1) & 0.6(0.1) \\
     & & & 6.7(1.9) & -39.4(0.1) & 0.6(0.2) \\ 
     & & & 4.3(1.5) & -40.1(0.2) & 2.1(0.4) \\ 
    $335.060\!-\!0.427^{a}$ & 16:29:23.146 & -49:12:27.34 &  2.9(0.5) & 
     -39.3(0.1) & 2.9(0.4) \\
     & & & 1.7(0.8) & -39.2(0.1) & 0.4(0.3) \\
    $25.826\!-\!0.178$      & 18:39:03.621 & -06:24:09.29 &  53.8(1.0) & 
     90.2(0.05) & 0.4(0.05) \\
     & & & 15.4(1.0) & 98.4(0.05) & 0.7(0.05)  \\
     & & & 10.0(1.0) & 91.7(0.05) & 0.3(0.05)  \\
     & & & 9.8(0.4) & 94.2(0.1) & 4.3(0.3)  \\
     & & & 4.6(0.7) & 91.4(0.1) & 2.5(0.3)  \\
     & & & 1.7(0.9) & 98.5(0.4) & 1.4(0.9)  \\
    $26.602\!-\!0.220$      & 18:40:38.550 & -05:43:56.20 &  8.3(1.6) &
     108.0(0.05) & 0.3(0.1) \\
     & & & 2.8(0.7) & 106.8(0.2) & 1.0(0.6) \\ \hline
  \end{tabular}
  \label{tab:det95}
\end{table*}

\begin{table*}
  \contcaption{}
  \begin{tabular}{lllrrrl} \hline
    {\bf Name} & {\bf Right Ascension} & {\bf Declination} & 
    {\bf Peak Flux} & {\bf Velocity} & {\bf Full Width}   & {\bf Ref}     \\
               & {\bf (J2000)}         & {\bf (J2000)}     & 
    {\bf Density}   & {\bf (\kms\/)} & {\bf Half Maximum} & \\ 
               & {\bf (h~m~s)}      & {\bf (\degrees\/ \arcmin\/ \arcsec\/)} &
    {\bf (Jy)}      &                & {\bf (\kms\/)} \\ [2mm] \hline
    $27.286\!+\!0.151^{a}$  & 18:40:34.476 & -04:57:13.44 &  3.6(0.4) &
     31.6(0.05) & 0.8(0.1) & * \\ 
     & & & 3.1(0.5) & 33.1(0.05) & 0.5(0.1) & \\
    $27.369\!-\!0.164$      & 18:41:50.982 & -05:01:28.23 &  22.9(0.8) &
     95.3(0.05)& 0.5(0.05) & * \\
     & & & 20.2(0.9) & 93.7(0.1) & 1.8(0.1) & \\
     & & & 13.7(1.4) & 94.2(0.05) & 0.5(0.1) & \\
     & & & 6.2(0.4) & 92.0(0.1) & 7.0(0.3) & \\
     & & & 5.8(0.6) & 90.6(0.05) & 0.6(0.1) & \\
     & & & 4.1(0.7) & 92.1(0.05) & 0.4(0.1) & \\
    $28.303\!-\!0.389$      & 18:44:22.056 & -04:17:48.84 & 4.7(1.0) &
     74.0(0.05) & 0.2(0.1) & * \\
    $29.907\!-\!0.040$      & 18:46:03.585 & -02:42:36.34 & 26.6(1.0) &
     98.2(0.05) & 0.3(0.05) & * \\
    $29.974\!-\!0.029$      & 18:46:08.505 & -02:38:42.66 & 3.1(0.6) &
     96.4(0.1) & 0.5(0.2) & 1 \\ 
     & & & 2.2(0.8) & 97.0(0.1) & 0.4(0.2) & \\ \hline 
  \end{tabular}
\end{table*}

\begin{figure*}
  \specdfig{comb_326_475}{comb_326_641}
  \specdfig{comb_326_859}{comb_327_392}
  \specdfig{comb_327_618}{comb_328_237}
  \caption{Spectra of the 95.1-GHz class~I methanol maser sources (top) and 
    the corresponding 6.6-GHz class~II methanol masers (bottom).  The
    95.1-GHz spectra are from the current work, except for those
    sources where \citet{VES+00} is the only reference listed in
    Table~\ref{tab:comparison}.  The 6.6-GHz spectra are from either
    \citet{EVM+96}, \citet{E96} or the current work, as listed in the
    relevant column of Table~\ref{tab:comparison}. All 95.1-GHz
    spectra have been Hanning smoothed.}
  \label{fig:meth95}
\end{figure*}

\begin{figure*}
  \specdfig{comb_328_809}{comb_329_029}
  \specdfig{comb_329_031}{comb_329_066}
  \specdfig{comb_329_183}{comb_329_469}
  \specdfig{comb_331_132}{comb_331_342}
  \contcaption{}
\end{figure*}

\begin{figure*}
  \specdfig{comb_331_442}{comb_332_295}
  \specdfig{comb_332_604}{comb_332_942}
  \specdfig{comb_332_963}{95_333_029}
  \specdfig{comb_333_121}{comb_333_128}
  \contcaption{}
\end{figure*}

\begin{figure*}
  \specdfig{comb_333_130}{comb_333_163}
  \specdfig{comb_333_184}{comb_333_234}
  \specdfig{comb_333_315}{comb_333_466}
  \specdfig{comb_333_562}{comb_335_060}
  \contcaption{}
\end{figure*}

\begin{figure*}
  \specdfig{comb_25_826}{comb_26_602}
  \specdfig{comb_27_286}{comb_27_369}
  \specdfig{comb_28_303}{comb_29_907}
  \specsfig{comb_29_974}
  \contcaption{}
\end{figure*}

\begin{table*}
  \caption{Class~II methanol maser sites not detected in the 95.1-GHz class I 
    transition.  The RMS noise level listed is that measured in the Hanning
    smoothed spectra.  Notes : $^{a}$ sources not within the statistically 
    complete sample of 6.6-GHz methanol masers, $^{b}$ emission detected 
    towards this source, but it appears to be a sidelobe response to 
    29.907-0.040.}  
  \begin{tabular}{lllcr} \hline
    {\bf Name} & {\bf R.A.(J2000)} & {\bf Dec.(J2000)}                      & 
    {\bf Velocity Range} & {\bf 3$\sigma$} \\
               & {\bf (h~m~s)}     & {\bf (\degrees\/ \arcmin\/ \arcsec\/)} & 
    {\bf (\kms\/)}       & {\bf (Jy)}      \\ [2mm] \hline
    $327.120\!+\!0.511$     & 15:47:32.729 & -53:52:38.90 &  -120 -- -55  & 4.4 \\
    $327.401\!+\!0.445$     & 15:49:19.4   & -53:45:14    &  -122 -- -46  & 4.2 \\
    $327.402\!+\!0.444$     & 15:49:19.523 & -53:45:14.21 &  -122 -- -46  & 4.2 \\
    $327.590\!-\!0.094$     & 15:52:36.824 & -54:03:18.97 &  -122 -- -46  & 4.8 \\
    $329.407\!-\!0.459$     & 16:03:32.662 & -53:09:26.98 &  -105 -- -29  & 4.2 \\
    $330.952\!-\!0.182^{a}$ & 16:09:52.372 & -51:54:57.89 &  -127 -- -50  & 4.2 \\
    $331.278\!-\!0.188$     & 16:11:26.596 & -51:41:56.67 &  -118 -- -40  & 3.8 \\
    $331.542\!-\!0.066$     & 16:12:09.020 & -51:25:47.60 &  -124 -- -47  & 3.8 \\
    $331.556\!-\!0.121^{a}$ & 16:12:27.210 & -51:27:38.20 &  -143 -- -66  & 4.1 \\
    $333.029\!-\!0.015$     & 16:18:44.167 & -50:21:50.77 &   -92 -- -16  & 3.0 \\
    $333.068\!-\!0.447$     & 16:20:48.995 & -50:38:40.72 &   -92 -- -16  & 3.4 \\
    $333.646\!+\!0.058^{a}$ & 16:21:09.140 & -49:52:45.90 &  -127 -- -50  & 3.8 \\ 
    $333.683\!-\!0.437$     & 16:23:29.794 & -50:12:08.69 &   -45 --  32  & 3.3 \\
    $333.931\!-\!0.135$     & 16:23:14.831 & -49:48:48.87 &   -76 --   0  & 4.1 \\
    $334.635\!-\!0.015$     & 16:25:45.729 & -49:13:37.51 &   -70 --   7  & 3.2 \\
    $334.935\!-\!0.098^{a}$ & 16:27:24.250 & -49:04:11.30 &   -67 --  20  & 3.0 \\ 
    $25.411\!+\!0.105$      & 18:37:16.918 & -06:38:28.23 &    58 -- 136  & 4.2 \\
    $25.386\!+\!0.005$      & 18:37:35.522 & -06:42:34.34 &    56 -- 132  & 3.9 \\
    $25.53\!+\!0.38$        & 18:36:32.6   & -06:24:26    &    52 -- 130  & 3.7 \\
    $25.710\!+\!0.044$      & 18:38:03.097 & -06:24:14.31 &    56 -- 132  & 4.1 \\
    $26.528\!-\!0.266$      & 18:40:40.227 & -05:49:07.67 &    66 -- 142  & 4.0 \\
    $27.223\!+\!0.137$      & 18:40:30.434 & -05:00:58.88 &    80 -- 156  & 3.0 \\
    $28.151\!-\!0.002$      & 18:42:42.494 & -04:15:18.34 &    62 -- 138  & 3.1 \\
    $28.829\!+\!0.488$      & 18:42:12.433 & -03:25:39.56 &    46 -- 122  & 3.2 \\
    $28.863\!-\!0.237$      & 18:44:51.000 & -03:43:44.17 &    50 -- 126  & 3.3 \\
    $28.810\!+\!0.360$      & 18:42:37.485 & -03:30:12.52 &    56 -- 130  & 3.4 \\
    $29.313\!-\!0.165$      & 18:45:24.974 & -03:17:44.47 &     6 --  82  & 3.4 \\
    $29.867\!-\!0.042$      & 18:45:59.530 & -02:44:47.23 &    56 -- 132  & 3.5 \\
    $29.865\!-\!0.007$      & 18:45:51.765 & -02:43:57.38 &    52 -- 128  & 3.5 \\
    $29.895\!-\!0.047^{b}$  & 18:46:03.653 & -02:43:24.89 &    60 -- 136  & 3.1 \\
    $29.918\!-\!0.035^{b}$  & 18:46:03.690 & -02:41:52.53 &    60 -- 136  & 2.9 \\
    $29.923\!+\!0.059$      & 18:45:44.184 & -02:39:03.71 &    60 -- 136  & 3.1 \\
    $30.009\!-\!0.017$      & 18:46:09.851 & -02:36:30.75 &    60 -- 136  & 3.4 \\ \hline
  \end{tabular}
  \label{tab:nondet95}
\end{table*}

\begin{figure}
  \psfig{file=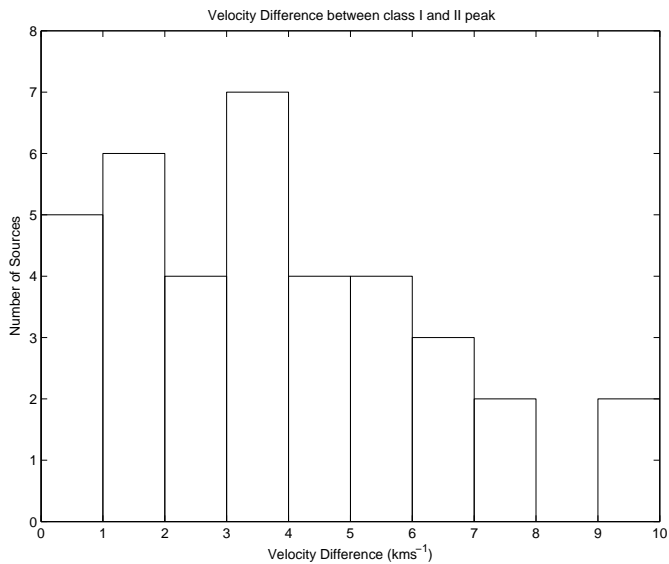,height=0.425\textwidth}
  \caption{The difference between the peak velocity of the class~I 95.1-GHz 
    methanol masers and the corresponding class~II 6.6-GHz transition}
  \label{fig:peakdiff}
\end{figure}

\begin{figure}
  \psfig{file=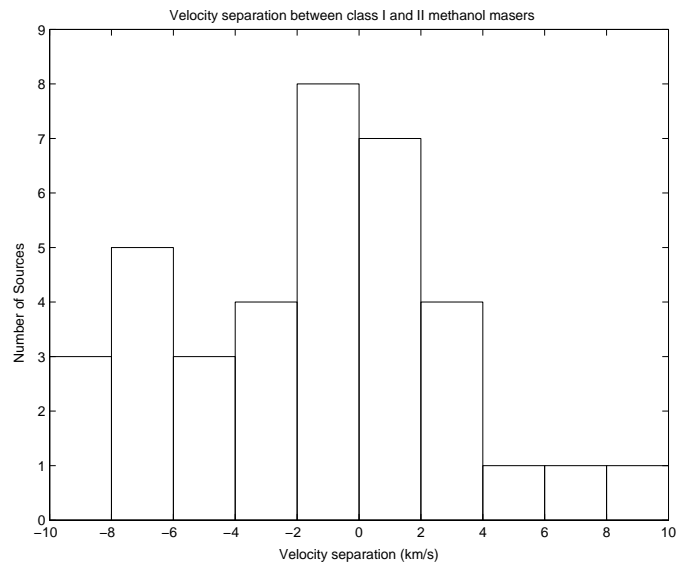,height=0.425\textwidth}
  \caption{The difference between the velocity ranges of the class~I 95.1-GHz 
    methanol masers and the corresponding class~II 6.6-GHz transition.
    A negative value indicates that the emission from the two classes
    overlaps.}
  \label{fig:separation}
\end{figure}

The primary purpose of these observations was to realise a search for
95.1-GHz class~I methanol masers towards a statistically complete
sample of 6.6-GHz class~II methanol masers.
Table~\ref{tab:comparison} gives the intensity and velocity of the
peak flux density and the velocity range for both the class~I and
class~II methanol masers for all the 6.6-GHz masers detected in the Mt
Pleasant survey (not just those in the statistically complete sample)
and the additional sources listed in section~\ref{sec:obs}.  There are
two 6.6-GHz class~II methanol maser sources from the statistically
complete sample which were accidentally omitted from the sources
searched for 95.1-GHz class~I masers ($328.254\!-\!0.254$ \&
$28.201\!-\!0.049$).  The data for the class~II emission in these
sources is included in Table~\ref{tab:comparison}.  Fortunately the
omission of two of sixty-eight sources does not significantly effect
the reported statistics or conclusions of this study.  For sources
where both classes of methanol maser were detected spectra of each are
shown on a common velocity scale in Figure~\ref{fig:meth95}.  Many of
the class~II methanol maser spectra contain emission from multiple
sources, present within a single Mt Pleasant beam (7 arcmin at
6.6~GHz).  Where this occurs the emission from the nearby sources is
indicated in the spectra.  Examination of Fig.~\ref{fig:meth95} shows
that the peaks of the two classes of methanol masers essentially never
coincide in velocity.  The median difference between the peak velocity
of the class~I and class~II transitions is 3.6~\kms\ 
(Fig.~\ref{fig:peakdiff}).  This is less than the median velocity
width (5~\kms) for the class~II methanol masers with associated
class~I emission, suggesting a significant degree of overlap exists
between the velocity ranges of the two classes. This can be examined
directly by investigating the minimum separation of the velocity
ranges between the class~I and II emission
(Figure~\ref{fig:separation}).  The separation here is defined to be
the difference between the lowest velocity of the class with the
larger mean velocity, and the highest velocity of the class with the
lower mean velocity.  For example for $326.475\!+\!0.703$ the class~I
masers cover a velocity range from $-43$ -- $-38$~\kms\ (a mean
velocity of $-40.5$~\kms), while the class~II masers have a velocity
range from $-51$ -- $-37$~\kms\ (a mean velocity of $-44$~\kms), so
the class~I masers have the higher mean velocity and the separation is
$-43 - -37 = -6$~\kms.  A negative value in Fig.~\ref{fig:separation}
indicates overlapping velocity ranges and a positive value,
non-overlapping velocity ranges.  The median separation observed in
the sample is -2~\kms, i.e. more than 50 percent of sources show an
overlap in the velocity ranges of the two classes.  This is contrary
to the suggestion of \citet{SKVO94} who claim an anti-correlation
between the velocities of the two classes.  Fig.~\ref{fig:separation}
shows that in the majority of cases (23 of 37) there is an overlap in
the velocity ranges and sometimes the overlap is large.

Figure~\ref{fig:meth95} shows that there are significant differences
in the typical spectral profiles of the two transitions.  The class~II
emission in essentially all the sources consists of one or more narrow
($<$ 1~\kms) spectral features.  In contrast the class~I emission
often contains only a single narrow peak, apparently superimposed on
broader emission features.  Some class~I sources do have multiple
narrow spectral features, but these are the exception rather than the
rule and essentially all sources have the broad emission features
rarely seen in the class~II emission.  Table~\ref{tab:det95} gives
details of the Gaussian profile fitting for each of the class~I maser
sources and shows that spectral features with widths $>$ 1~\kms\ are
ubiquitous. It is not clear whether the broad features are
quasi-thermal/quasi-maser emission, or perhaps due to blending of a
number of weaker maser features.  The degree of symmetry and lack of
multiple peaks favours the former interpretation.  However,
interferometric observations are required to provide a definitive
answer.  The peak flux density and the noise level in the observations
of the two classes are comparable and so it cannot be attributed to
poorer signal to noise ratio in the class~I observations.

\begin{table*}
  \caption{Comparison of the peak flux density and velocity range of 6.6-GHz 
    class II and 95.1-GHz class I methanol masers.  Notes : $^{a}$ sources 
    not within the statistically complete sample of 6.6-GHz methanol masers.
    References from which the 6.6- and 95.1-GHz information was obtained : 
    * = this paper ;
    1 = \citet{VES+00} ;
    2 = \citet{E96} ;
    3 = \citet{EVM+96};
    4 = \citet{C96}.
    For a comprehensive list of references for each 6.6-GHz masers see 
    \citet{PMB04}.}
  \begin{tabular}{lrrrlrrrl} \hline
    {\bf Source} & \multicolumn{4}{c}{\bf 95.1-GHz Class I methanol masers} & 
    \multicolumn{4}{c}{\bf 6.6-GHz Class II methanol masers} \\
    {\bf Name}   & \multicolumn{2}{c}{\bf Peak Flux } & {\bf Velocity} & 
    {\bf Ref} & \multicolumn{2}{c}{\bf Peak Flux} & {\bf Velocity} & 
    {\bf Ref} \\
               & {\bf Density} & {\bf Velocity} & {\bf Range} &          &
    {\bf Density} & {\bf Velocity} & {\bf Range}         & \\ 
               & {\bf (Jy)}     & {\bf (\kms)}   & {\bf (\kms)} &        & 
    {\bf (Jy)}     & {\bf (\kms)}   & {\bf (\kms)} &       \\ \hline
    $285.32\!-\!0.03$       &     &        &            & 1      &   11 &    0.6 & -8 -- 3    &  2         \\
    $291.28\!-\!0.71$       &     &        &            & 1      &   70 &  -29.7 & -31 -- -26 &  2         \\
    $293.84\!-\!0.78$       &     &        &            & 1      &    4 &   36.9 & 36 -- 39   &  2         \\
    $293.95\!-\!0.91$       &     &        &            & 1      &    8 &   41.4 &         &     2         \\
    $326.475\!+\!0.703$     &  29 &  -41.0 & -43 -- -38 & 1      &  109 &  -38.1 & -51 -- -37 &  *         \\
    $326.641\!+\!0.613^{a}$ &  19 &  -39.9 & -41 -- -37 & 1      &   17 &  -43.1 & -44 -- -40 &  3         \\
    $326.662\!+\!0.521$     &     &        &            & 1      &    5 &  -41.0 & -42 -- -29 &  3         \\
    $326.859\!-\!0.677^{a}$ &  11 &  -67.0 & -68 -- -66 & 1      &   10 &  -57.6 & -59 -- -57 &  3         \\
    $327.120\!+\!0.511$     &     &        &            & *      &   80 &  -87.1 & -90 -- -83 &  3         \\
    $327.392\!+\!0.199$     &   8 &  -89.5 & -90 -- -88 & 1      &    9 &  -84.6 & -86 -- -82 &  3         \\
    $327.401\!+\!0.445$     &     &        &            & *      &  106 &  -82.6 & -84 -- -75 &  3         \\
    $327.402\!+\!0.444$     &     &        &            & *      &  106 &  -82.6 & -84 -- -75 &  3         \\
    $327.590\!-\!0.094$     &     &        &            & *,1    &    3 &  -86.3 &         &     3         \\
    $327.618\!-\!0.111$     &   7 &  -88.4 & -89 -- -87 & 1      &    2 &  -97.5 &         &     3         \\
    $327.945\!-\!0.115$     &     &        &            & 1      &    7 &  -51.7 & -52 -- -51 &  3         \\
    $328.254\!-\!0.532$     &     &        &            &        &  425 &  -37.4 & -50 -- -36 &  3         \\
    $328.237\!-\!0.548^{a}$ &  10 &  -41.3 & -45 -- -38 & *,1    &  421 &  -44.9 & -46 -- -34 &  3         \\
    $328.809\!+\!0.633^{a}$ &  45 &  -40.6 & -46 -- -37 & *,1    &  278 &  -44.5 & -47 -- -43 &  3         \\
    $329.031\!-\!0.198$     &  20 &  -43.7 & -48 -- -39 & *      &   25 &  -41.9 & -47 -- -41 &  3         \\
    $329.029\!-\!0.205$     &  18 &  -43.6 & -47 -- -37 & 1      &  275 &  -37.5 & -41 -- -34 &  3         \\
    $329.066\!-\!0.308$     & 5.6 &  -42.1 & -43 -- -40 & *,1    &   24 &  -43.9 & -48 -- -43 &  3         \\
    $329.183\!-\!0.314$     &   8 &  -50.1 & -51 -- -47 & 1      &   13 &  -55.7 & -60 -- -51 &  3         \\
    $329.339\!+\!0.148$     &     &        &            & 1      &   14 & -106.5 &-107 -- -105&  3         \\
    $329.407\!-\!0.459$     &     &        &            & *      &  144 &  -66.8 & -71 -- -66 &  3         \\
    $329.469\!+\!0.502$     &   9 &  -66.6 & -70 -- -66 & 1      &   13 &  -72.1 & -73 -- -65 &  3         \\
    $329.622\!+\!0.138$     &     &        &            & 1      &   30 &  -60.1 & -69 -- -59 &  3         \\
    $329.610\!+\!0.114$     &     &        &            & 1      &   30 &  -60.1 & -69 -- -59 &  3         \\
    $330.952\!-\!0.182^{a}$ &     &        &            & *      &    7 &  -87.6 & -89 -- -87 &  3         \\
    $331.425\!+\!0.264^{a}$ &     &        &            & 1      &   25 &  -88.6 & -91 -- -88 &  3         \\
    $331.120\!-\!0.118^{a}$ &     &        &            & 1      &    9 &  -93.2 & -95 -- -88 &  4         \\
    $331.132\!-\!0.244$     &  32 &  -91.2 & -81 -- -92 & *,1    &   34 &  -84.4 & -92 -- -84 &  3         \\
    $331.278\!-\!0.188$     &     &        &            & *      &  165 &  -78.1 & -86 -- -78 &  3         \\
    $331.342\!-\!0.346$     &  13 &  -65.7 & -67 -- -65 & *,1    &   66 &  -67.4 & -68 -- -64 &  3         \\
    $331.442\!-\!0.187^{a}$ & 4.9 &  -91.7 & -92 -- -87 & *,1    &   70 &  -88.5 & -93 -- -84 &  3         \\
    $331.542\!-\!0.066$     &     &        &            & *      &   12 &  -84.1 & -87 -- -83 &  3         \\
    $331.556\!-\!0.121^{a}$ &     &        &            & *      &   35 & -103.4 & -105 -- -94&  3         \\
    $332.094\!-\!0.421$     &     &        &            & 1      &   16 &  -61.4 & -62 -- -58 &  3         \\
    $332.295\!-\!0.094$     & 5.9 &  -49.7 & -50 -- -48 & *,1    &    6 &  -47.0 & -47 -- -42 &  3         \\
    $332.351\!-\!0.436$     &     &        &            & 1      &    4 &  -53.1 &         &     3      \\
    $332.560\!-\!0.148$     &     &        &            & 1      &    5 &  -51.0 & -56 -- -49 &  3         \\
    $332.604\!-\!0.167$     &  10 &  -45.7 & -47 -- -45 & 1      &    5 &  -51.0 & -56 -- -49 &  3         \\
    $332.942\!-\!0.686^{a}$ &   9 &  -48.9 & -51 -- -47 & *      &   21 &  -52.9 & -54 -- -52 &  3         \\
    $332.963\!-\!0.679^{a}$ &  11 &  -47.4 & -51 -- -46 & *      &   41 &  -45.9 & -48 -- -38 &  3         \\
    $333.029\!-\!0.015$     &     &        &            & *      &    3 &  -53.6 & -61 -- -53 &  3         \\
    $333.029\!-\!0.063^{a}$ & 4.6 &  -40.9 & -42 -- -39 & *      &    4 &  -40.4 & -41 -- -40 &  4         \\
    $333.068\!-\!0.447$     &     &        &            & *      &   12 &  -54.5 & -55 -- -53 &  3         \\
    $333.121\!-\!0.434$     &  11 &  -50.7 & -56 -- -47 & *      &   11 &  -49.3 & -50 -- -48 &  3         \\
    $333.128\!-\!0.440$     &  46 &  -48.3 & -55 -- -45 & *,1    &    3 &  -44.4 & -45 -- -42 &  3         \\
    $333.130\!-\!0.560^{a}$ &  10 &  -59.0 & -61 -- -54 & *      &   17 &  -56.8 & -63 -- -52 &  3         \\
    $333.163\!-\!0.101^{a}$ & 4.6 &  -91.2 & -92 -- -90 & *      &    8 &  -95.3 & -95 -- -91 &  3         \\
    $333.184\!-\!0.091$     &   7 &  -87.0 & -87 -- -85 & *      &    7 &  -82.0 & -85 -- -81 &  3         \\
    $333.234\!-\!0.062$     & 150 &  -87.4 & -91 -- -84 & *,1    &    7 &  -84.7 & -85 -- -81 &  3         \\
    $333.315\!+\!0.105$     & 4.2 &  -44.5 & -50 -- -43 & *      &    9 &  -43.7 & -50 -- -41 &  3         \\
    $333.466\!-\!0.164$     &  12 &  -42.8 & -46 -- -40 & *      &   41 &  -42.4 & -49 -- -37 &  3         \\
    $333.562\!-\!0.025$     &  17 &  -39.8 & -41 -- -39 & *      &   39 &  -35.9 & -37 -- -34 &  3         \\
    $333.646\!+\!0.058^{a}$ &     &        &            & *      &    3 &  -87.3 & -89 -- -82 &  4         \\
    $333.683\!-\!0.437$     &     &        &            & *      &   19 &   -5.2 & -6 -- -4   &  3         \\
    $333.931\!-\!0.135$     &     &        &            & *      &    7 &  -36.8 & -37 -- -36 &  3         \\
    $334.635\!-\!0.015$     &     &        &            & *      &   61 &  -30.1 & -31 -- -27 &  3         \\ \hline
  \end{tabular}
  \label{tab:comparison}
\end{table*}

\begin{table*}
  \contcaption{}
  \begin{tabular}{lrrrlrrrl} \hline
    {\bf Source} & \multicolumn{4}{c}{\bf 95.1-GHz Class I methanol masers} & 
    \multicolumn{4}{c}{\bf 6.6-GHz Class II methanol masers} \\
    {\bf Name}   & \multicolumn{2}{c}{\bf Peak Flux} & {\bf Velocity} & 
    {\bf Ref} & \multicolumn{2}{c}{\bf Peak Flux} & {\bf Velocity} & 
    {\bf Ref} \\
               & {\bf Density} & {\bf Velocity} & {\bf Range} &          &
    {\bf Density} & {\bf Velocity} & {\bf Range}         & \\ 
               & {\bf (Jy)}     & {\bf (\kms)}   & {\bf (\kms)} &        & 
    {\bf (Jy)}     & {\bf (\kms)}   & {\bf (\kms)} &       \\ \hline
    $334.935\!-\!0.098^{a}$ &     &        &            & *      &    7 &  -19.5 & -22 -- -17 &  4         \\
    $335.060\!-\!0.427^{a}$ & 4.6 &  -39.1 & -41 -- -37 & *      &   15 &  -47.0 & -48 -- -39 &  3         \\ 
    $25.386\!+\!0.005$      &     &        &            & *      &    2 &   95.7 & 94 -- 98   &  2         \\
    $25.411\!+\!0.105$      &     &        &            & *,1    &    8 &   97.0 & 96 -- 98   &  2         \\
    $25.53+\!0\!.38 $       &     &        &            & *,1    &    6 &   95.6 & 89 -- 96   &  2         \\
    $25.710\!+\!0.044$      &     &        &            & *      &  502 &   95.7 & 89 -- 101  &  2         \\
    $25.826\!-\!0.178$      &  58 &   90.2 & 90 -- 99   & *,1    &   70 &   91.7 & 90 -- 100  &  2         \\
    $26.528\!-\!0.266$      &     &        &            & *      &    9 &  104.5 & 104 -- 105 &  2         \\
    $26.602\!-\!0.220$      &   8 &  108.1 & 108 -- 114 & *      &    9 &  103.8 & 102 -- 115 &  2         \\
    $27.223\!+\!0.137$      &     &        &            & *      &   22 &  117.7 & 110 -- 121 &  2         \\
    $27.286\!+\!0.151^{a}$  & 4.0 &   31.7 & 31 -- 34   & *      &   23 &   34.9 & 34 -- 36   &  *         \\
    $27.369\!-\!0.164$      &  35 &   94.2 & 88 -- 96   & *      &   42 &  100.5 & 88 -- 104  &  2         \\
    $28.151\!-\!0.002$      &     &        &            & *      &   37 &  101.2 & 100 -- 105 &  2         \\
    $28.201\!-\!0.049$      &     &        &            &        &    2 &   98.9 & 94 -- 99   &  2         \\
    $28.303\!-\!0.389$      & 4.3 &   73.9 & 74 -- 75   & *      &   71 &   81.1 & 80 -- 94   &  2         \\
    $28.829\!+\!0.488$      &     &        &            & *      &    5 &   83.2 & 83 -- 84   &  2         \\
    $28.863\!-\!0.237$      &     &        &            & *      &   65 &   83.5 & 81 -- 93   &  2         \\
    $28.810\!+\!0.360$      &     &        &            & *      &    6 &   91.1 & 87 -- 93   &  2         \\
    $29.313\!-\!0.165$      &     &        &            & *      &    4 &   48.8 & 43 -- 50   &  2         \\
    $29.867\!-\!0.042$      &     &        &            & *      &   35 &  101.4 & 100 -- 104 &  2         \\
    $29.865\!-\!0.007$      &     &        &            & *      &    2 &  103.4 & 100 -- 104 &  2         \\
    $29.895\!-\!0.047$      &     &        &            & *      &   40 &   96.8 & 96 -- 99   &  2         \\
    $29.907\!-\!0.040$      &  26 &   98.2 & 98 -- 99   & *      &    3 &   99.0 & 95 -- 99   &  2         \\
    $29.918\!-\!0.035$      &     &        &            & *      &   40 &   96.8 & 95 -- 99   &  2         \\
    $29.923\!+\!0.059$      &     &        &            & *      &    5 &   99.0 & 95 -- 102  &  2         \\
    $29.974\!-\!0.029$      & 3.4 &   96.3 & 96 -- 97   & *,1    &   80 &   96.0 & 95 -- 100  &  2         \\
    $30.009\!-\!0.017$      &     &        &            & *      &   10 &   98.3 & 98 -- 103  &  2         \\ \hline
  \end{tabular}
\end{table*}

Focusing on the statistically complete sample of sixty-six class~II
methanol masers (those masers not included in this sample are noted in
Tables~\ref{tab:det95} and \ref{tab:comparison}), there are a total of
twenty-five detections of associated class~I masers.  This represents
a detection rate of $38\pm6$ per cent for class~I methanol masers
towards class~II sources.  However, as the 95.1-GHz transition is not
the strongest of the class~I transition this figure should be taken as
a lower limit.  \citet{VES+00} showed that the 44-GHz transition is
typically a factor of 3 stronger than the 95.1-GHz transition and so
the number of class~II methanol masers with an associated class~I
maser is likely to be of the order of 50 per cent or more.

\subsection{Comments on individual sources} \label{sec:indiv}

{\em 326.475$\!+\!$0.703:} The class~II methanol maser emission in
this source is dominated by two strong peaks separated by more than
10~\kms\ each of which have a noticeably sharper inner edge than outer
edge.  The class~I maser emission lies in between the two class~II
peaks, with the red wing of the 95.1-GHz masers overlapping the
velocity range of the strongest 6.6-GHz emission.  The 6.6-GHz
observation spectrum in Fig.~\ref{fig:meth95} is a new observation,
however the relative intensity of the two main features appears to
have changed little in the 11 years since the discovery of this source
\citep*{VGM95}

\noindent
{\em 326.859$\!-\!$0.667:} The class~I and II methanol maser emission
in this source each have comparable peak flux density and velocity
range, but are offset from each other by approximately 10~\kms, one of
the largest offsets in the sample.

\noindent
{\em 328.237$\!-\!$0.548:} The class~II methanol maser emission in
this source covers two separate velocity ranges and the class~I
emission lies in between these.  95.1-GHz class~I methanol maser
emission was first observed in this source by \citet{VES+00} (who
called it $328.24\!-\!0.55$) and there has been no measurable change
in the peak flux density of the source in the 12 month period between
the two observations.

\noindent
{\em 328.809$\!+\!$0.633:} This strong class~II methanol maser source
is associated with a strong class~I methanol maser.  The peak flux
density of the 95.1-GHz methanol masers is more than a factor four
greater than observed by \citet{VES+00} a year earlier.  The bulk of
the class~I maser emission is red-shifted compared to the class~II
emission, but the wing overlaps the velocity of the strongest class~II
masers.

\noindent
{\em 329.029$\!-\!$0.205 \& 329.031$\!-\!$0.198:} These two class~II
maser sites are separated by 26~arcsec and each also has
associated 95.1-GHz class~I methanol maser emission.  The spectrum of the
95.1-GHz methanol masers in 329.029$\!-\!$0.205 are blue-shifted with
respect to the strongest 6.6-GHz class II emission.  The class~I
methanol maser emission in 329.031$\!-\!$0.198 partially overlaps that
of 329.029$\!-\!$0.205, but the peak velocity differs slightly.  In
contrast to 329.029$\!-\!$0.205, in 329.031$\!-\!$0.198 the velocity
ranges of the class~I and II methanol masers are largely overlapping.

\noindent
{\em 329.469$\!+\!$0.502:} This relatively weak, newly discovered
95.1-GHz class~I maser source shows weak emission at velocities
between the two class~II maser velocity ranges in this source.

\noindent
{\em 331.132$\!-\!$0.244:} The 95.1-GHz class~I masers in this source
differ from the majority in showing multiple narrow peaks over a
substantial velocity range.  The velocity of the class~I and II
methanol masers in this source overlap to a high degree.  The
strongest class~I emission is at the opposite end of the velocity
range to the strongest class~II masers.  However, the 6.6-GHz methanol
masers in this sources are known to be highly variable \citep*{CVE95}.
The intensity of the entire 95.1-GHz spectrum has increased compared
to that observed by of \citet{VES+00}, suggesting that it is also
variable.

\noindent
{\em 331.342$\!-\!$0.346:} The 95.1-GHz class~I maser in this source
contains a single narrow peak, lacking the broader velocity feature
seen in many sources.  The class~I maser peak lies in the middle of
the class~II maser velocity range.

\noindent
{\em 332.942$\!-\!$0.686 \& 332.963$\!-\!$0.679:} The two regions of
6.6-GHz class~II methanol maser emission are separated by 1.3~arcmin,
but have non-overlapping velocity ranges.  Each region also has newly
discovered, class~I maser emission, the velocity ranges of which
overlap, but have distinct peak velocities.  The class~I emission in
each case lies in between the class~II velocity ranges of the two
sources.

\noindent
{\em 333.029$\!-\!$0.063:} The new 6.6-GHz Mt Pleasant observations of
this source failed to detect any class~II methanol maser emission
stronger than approximately 2~Jy, so only the 95-GHz spectrum is shown
for this source.  The observations of \citet{C96} show that the
velocity of the peak of the class~I and II methanol masers are
separated by 0.5~\kms.

\noindent
{\em 333.121$\!-\!$0.434 \& 333.128$\!-\!$0.440:} There are a number
of class~II methanol maser sources in this region, two of which also
have associated class~I maser emission.  333.121$\!-\!$0.434 is a
newly discovered 95.1-GHz class~I maser source whose velocity range
overlaps the class~II emission.  The 95.1-GHz emission in
333.128$\!-\!$0.440 was also observed by \citet{VES+00}, but with a
significantly lower peak flux density.

\noindent
{\em 333.130$\!-\!$0.560:} The 6.6-GHz methanol maser in this source
has an interesting velocity profile consisting of three narrow peaks,
the strongest which is bracketed by approximately equal, weaker
features.  Intriguingly the newly discovered class~I 95.1-GHz masers
are centred on the same velocity, with symmetrical emission lying
within velocity range of the class~II methanol masers.  This source
would be an interesting candidate for high-resolution observations in
both transitions to determine the spatial relationship between the two
classes.

\noindent
{\em 333.163$\!-\!$0.101:} The 95.1-GHz methanol maser emission in
this source shows two weak peaks separated in velocity by nearly
20~\kms, one of which is close to the velocity of the class~II masers.
This velocity range is much greater than that seen in any of the other
class~I methanol masers in this sample, suggesting that perhaps it is
a separated source detected within the same beam.  This region of the
Galactic Plane has been covered in two independent untargeted
searches for 6.6-GHz class~II masers \citep{EVM+96,C96}.  If the
feature at -110~\kms\ is indeed an independent source detected by
serendipity it suggests there may be a substantial number of class~I
maser sources associated neither with class~II methanol masers, nor
high-mass star formation regions known from other tracers.

\noindent
{\em 333.234$\!-\!$0.062} With a peak flux density of 150~Jy the
95.1-GHz class~I methanol maser in this source is the strongest in
this sample and is one of the strongest class~I masers in this
transition. \citet{VES+00} observed a significantly lower peak flux
density (26~Jy) suggesting either a significant pointing offset in
their observations, or variability.  The class~I maser emission is
associated with a weak class~II methanol maser which has a single peak
and non-overlapping velocity range.

\noindent
{\em 333.315$\!+\!$0.105:} This weak, newly discovered 95.1-GHz
methanol maser has a velocity range which overlaps completely with the
class~II maser emission.

\noindent
{\em 333.466$\!-\!$0.164:} This newly discovered class~I methanol
maser source is unusual in that the peak velocity of the class~I and
II transitions differ by less than 0.5~\kms.

\noindent
{\em 25.826$\!-\!$0.178:} This site of this strong, newly discovered
95.1-GHz methanol maser was previously searched by \citet{VES+00}, but
they report no emission stronger than 3~Jy.  There is a difference in
the positions observed in this work and by \citet{VES+00} of
30~arcseconds, which is significant, but does not completely account
for the non-detection.  The class~I and II methanol maser emission in
this source is notable for having very similar spectra, in terms of
peak intensity, velocity range and general shape.

\noindent
{\em 27.286$\!+\!$0.151:} The velocity of the maser emission in this
source lies outside the range of the original Mt Pleasant survey.
However, this source was detected in ATCA follow-up observations of
27.223$\!+\!$0.137.  It was also independently discovered by
\citet{SVK+99} and \citet{SHK00}.  The 6.6-GHz spectrum shown in
Fig.~\ref{fig:meth95} is a new observation, the earlier observations
each show a lower peak flux density, but otherwise similar spectra.
The weak, newly detected class~I methanol maser in this source doesn't
overlap in velocity with the class~II emission.

\noindent
{\em 28.303$\!-\!$0.389:} The marginal detection of 95.1-GHz methanol
maser emission in this source requires follow-up to confirm if it is
real.  It is offset by more than 5~\kms\ from the class~II emission,
which is unusual, but the class~II emission does extend over a larger
than usual range in this source.

\noindent
{\em 29.907$\!-\!$0.040:} There are five additional class~II methanol
maser sites within a relatively small distance of this source and they
have overlapping velocity ranges.  A 95.1-GHz class~I methanol maser
was detected towards a number of the class~II sites.  However, it
appears to be a single source which is strongest at this location.
Disentangling the relationship of the different star-formation sites
within this complex will require interferometric observations.

\noindent
{\em 29.974$\!-\!$0.029:} The marginal detection of 95.1-GHz methanol
maser emission in this source requires a follow-up to confirm if it is
real.  Emission of comparable strength at the same velocity was
detected by \citet{VES+00} and so this emission is likely to be real.

\section{Discussion} \label{sec:discussion}

\begin{figure}
  \psfig{file=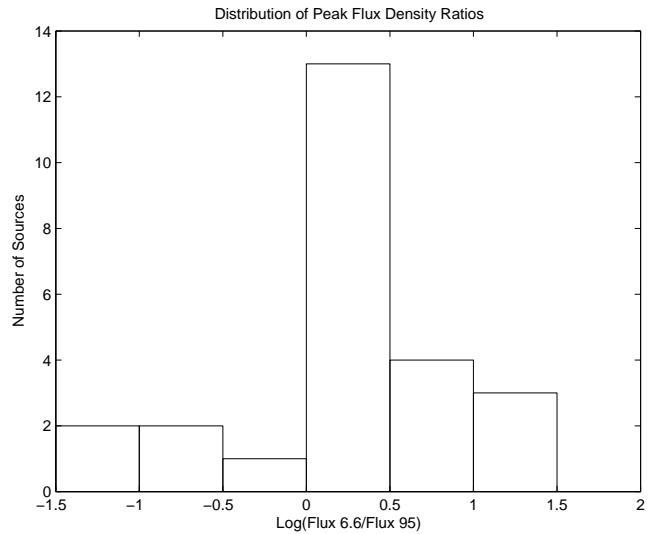,height=0.425\textwidth}
  \caption{Comparison of the distribution of ratio of the 6.6-GHz peak
    flux density to the associated 95-GHz peak flux density for the
    twenty-five methanol masers in the statistically complete sample.}
  \label{fig:ratio}
\end{figure}

\citet{SKVO94} claim that there is an anti correlation between the
flux density of class~I and class~II methanol masers within the same
region.  It is certainly true that the strongest class~II masers
typically do not have associated class~I masers (e.g. W3(OH),
NGC6334F) and vice versa.  This was one one of the factors in the
empirical classes as they were originally defined.  However, the
current observations show that for typical methanol masers there is no
anti-correlation between the peak flux density of the class~I and II
maser transitions.  Considering the twenty-five masers within the
statistically complete sample that have an associated class~I maser,
the 6.6- and 95.1-GHz peak flux densities are in general comparable
with a median ratio of 1.25.  Figure~\ref{fig:ratio} shows that the
distribution of peak flux density ratios is strongly peaked at values
slightly greater than one.

There are significant differences between the typical morphologies
observed for class~I and class~II methanol masers.  Class~II methanol
maser emission occurs in clusters with a maximum size of
30~milli-parsecs \cite{C97,PNEM98}.  The majority of sources have just
one cluster, some have a second cluster separated by a few arcseconds,
but more than two separate clusters in one region is very rare.  In
contrast the emission from class~I masers is typically spread over
much larger angular scales \cite{KS98,KHA04}.  This suggests enhanced
methanol abundance over a significant fraction of the star formation
region, but that conditions suitable for class~II masers are
relatively rare, while those that favour class~I masers are more
common.

Early interferometric observations of class~I methanol masers
suggested that they may be associated with outflows from high-mass
star formation regions, perhaps at the interface between the outflow
and the molecular cloud \citep{PM90,JGS+92}.  That they are observed
offset from \ionhy\ regions, and are collisionally pumped is also
consistent with this hypothesis.  Further support has recently been
provided by the observations of \citet{KHA04} who found a number of
sources for which there is a good correspondence between the 44~GHz
class~I masers and the molecular shock tracers H$_{2}$ and SiO.

It is well established that many water masers in star formation
regions are associated with outflows, some of which have velocities in
excess of 100~km/s and are highly collimated.  Interferometric
observations of water masers have shown that those with the largest
peak flux densities lie in outflows directed at angles close to the
plane of the sky \citep{GRM+81,GDS+81}.  With this geometry our
line-of-sight looks along the shock-front, giving a long velocity
coherent path and hence strong masers.  In contrast, masers which are
significantly offset from the systemic velocity of the system are in
outflows directed close to the line-of-sight, we view the shock-fronts
close to face on resulting in short gain paths and weak masers.

\begin{figure}
  \psfig{file=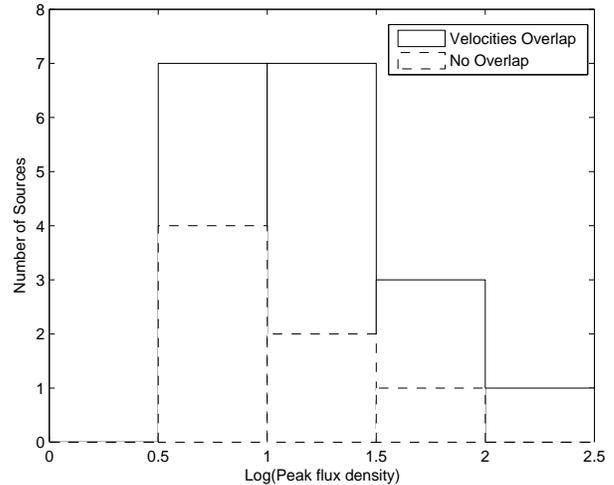,height=0.40\textwidth}
  \caption{Comparison of the distribution of peak flux density for class~I 
    methanol masers which have a velocity range which overlaps the
    class~II masers (solid line) and those which do not (dashed line).
    Only class~I masers associated with class~II masers in the
    statistically complete sample are included in this figure.}
  \label{fig:fcompare}
\end{figure}

The class~I methanol masers are clearly not associated with the same
outflows (or at least with the same parts of the outflows), as water
masers, as they are always within 10-20~\kms\ of the velocity of
class~II masers and thermal emissions.  However, we might expect
class~I methanol masers which are close to the systemic velocity of
the region to be associated with outflows directed close to the plane
of the sky and, analogous to water masers to be stronger.  If we
assume that the velocity of the class~II masers is close to the
systemic velocity of the region (which is generally the case), then we
can test this hypothesis by comparing the flux density of class~I
masers which overlap the velocity range of the class~II with those
that don't.  Figure~\ref{fig:fcompare} shows two histograms which make
this comparison.  Only those class~I methanol masers associated with
class~II masers in the statistically complete sample are included in
Fig.~\ref{fig:fcompare}.  The two samples (18 sources which have
overlap and 7 sources which don't) are too small to allow definitive
conclusions to be drawn.  However, there may be a higher percentage of
stronger class~I methanol masers associated with sources where there
is a velocity overlap between the two classes.

Two types of observations suggest that class~II methanol masers are
associated with the early stages of high-mass star formation.
\begin{enumerate}
\item Many class~II methanol masers are not associated with an UC\ionhy\ 
  region \citep{PNEM98,WBHR98}.
\item Class~II masers are associated with submillimetre and millimetre
  dust continuum emission which have SED consistent with deeply embedded
  high-mass protostellar objects \cite*{PHB02,WMABL03,MBH+04}
\end{enumerate}
The current star formation paradigm for low-mass star formation has
molecular outflows associated with the earliest stages of the process,
the so-called class~0 and class~I young stellar objects \citep{BT99}.
From this we might hypothesise that class~I methanol masers may be
associated with an even earlier stage of high-mass star formation than
is the general case for class~II masers.  Considering only methanol
masers we might expect sources with only class~I methanol masers to be
the youngest, with those which have both class~I and II masers being
at an intermediate phase and sources with only class~II methanol
masers being the most evolved.  We can test this hypothesis by looking
at the properties of the infrared sources and other maser species
associated with the methanol masers and seeing if there is any
difference between the class~II masers with and without an associated
class~I maser.  If a difference can be found then it would also
provide a method of targeting class~I methanol maser searches.

The two most common masers associated with high-mass star formation
(apart from class~II methanol masers) are the 1665-MHz transition of
OH and the 22-GHz transition of water.  An untargeted search of the
southern Galactic plane for main-line OH masers was made with the
Parkes telescope in the early 1980s \citep{CHG80,CH83a,CH87}, with
some regions more recently searched again at higher sensitivity with
the Australia Telescope Compact Array \citep{C98}.  The Parkes search
was sensitive to OH masers with a peak flux density in excess of 1~Jy
in the main-line transition at the epoch of the observations.  To date
there have been no blind searches for southern water masers.  However,
a search targeted towards the 6.6-GHz methanol masers detected in the
Mt Pleasant survey has been undertaken by \citet{H97}.
Table~\ref{tab:assoc} summarises the association of the class~II
methanol masers from the Mt Pleasant survey with 1665-MHz OH and
22-GHz water masers.  Considering only the class~II methanol masers in
the statistically complete sample, approximately 30~percent have an
associated 1665-MHz OH maser and approximately 40~percent have an
associated water maser.  However, there is no statistically
significant correlation, or anti-correlation between the presence of
OH or water masers and class~I methanol masers.

\begin{table}
  \caption{Associations of the class~II methanol masers with other maser
    species and {\em MSX} 21-\micron\ flux.  The {\em MSX} flux is measured
    directly from the images, for non-detections the flux density at the 
    maser location is given as an upper limit.  Notes : $^{a}$ sources 
    not within the statistically complete sample of 6.6-GHz methanol masers.
    References 1 = \citet{CH87} ; 2 = \citet{C98} ; 3 = \citet{CH83b} ;
    4 = \citet{CVEWN95} ; 5 = \citet{H97}}
  \begin{tabular}{lccccc} \hline
    {\bf Source} & \multicolumn{2}{c}{\bf 1665-MHz} & 
    \multicolumn{2}{c}{\bf 22-GHz} &
    \multicolumn{1}{c}{\bf MSX} \\
    {\bf Name}   & \multicolumn{2}{c}{\bf OH maser} & 
    \multicolumn{2}{c}{\bf \water\ maser} & 
    \multicolumn{1}{c}{\bf 21-\micron\ flux} \\
                 & {\bf Assoc} & {\bf Ref} & {\bf Assoc} & 
    {\bf Ref} & {\bf ($\mu$Wm$^{-2}$sr$^{-1}$)} \\ \hline
    $285.32\!-\!0.03$       & N  & 1 & Y & 5 & $<$1.25 \\
    $291.28\!-\!0.71$       & N  & 1 & Y & 5 & 1600 \\
    $293.84\!-\!0.78$       & N  & 1 & Y & 5 & $<$0.15 \\
    $293.95\!-\!0.91$       & N  & 1 & Y & 5 & $<$0.77 \\
    $326.475\!+\!0.703$     & N  & 2 & Y & 5 & $<$0.87 \\
    $326.641\!+\!0.613^{a}$ & N  & 2 & Y & 5 & $<$12.5 \\
    $326.662\!+\!0.521$     & N  & 2 & Y & 5 & 330 \\
    $326.859\!-\!0.677^{a}$ & N  & 2 & N & 5 & $<$3.6 \\
    $327.120\!+\!0.511$     & Y  & 2 & Y & 5 & 71  \\
    $327.392\!+\!0.199$     & N  & 2 & N & 5 & 9.6 \\
    $327.401\!+\!0.445$     & N  & 2 & N & 5 & 55 \\
    $327.402\!+\!0.444$     & Y  & 2 & Y & 5 & 41 \\
    $327.590\!-\!0.094$     & N  & 2 & N & 5 & 9.5 \\
    $327.618\!-\!0.111$     & N  & 2 & N & 5 & 8.1 \\
    $327.945\!-\!0.115$     & N  & 2 & N & 5 & 57 \\
    $328.254\!-\!0.532$     & Y  & 2 & Y & 5 & 53 \\
    $328.237\!-\!0.548^{a}$ & Y  & 2 & Y & 5 & 7.6 \\
    $328.809\!+\!0.633^{a}$ & Y  & 2 & N & 5 & 510 \\
    $329.031\!-\!0.198$     & Y  & 2 & Y & 5 & $<$4.0 \\
    $329.029\!-\!0.205$     & Y  & 2 & Y & 5 & $<$3.6 \\
    $329.066\!-\!0.308$     & Y  & 2 & N & 5 & 20 \\
    $329.183\!-\!0.314$     & Y  & 2 & Y & 5 & $<$5.7 \\
    $329.339\!+\!0.148$     & N  & 2 & N & 5 & 700 \\
    $329.407\!-\!0.459$     & Y  & 2 & Y & 5 & 19 \\
    $329.469\!+\!0.502$     & N  & 2 & N & 5 & $<$4.7 \\
    $329.622\!+\!0.138$     & N  & 2 & Y & 5 & $<$0.53 \\
    $329.610\!+\!0.114$     & N  & 2 & N & 5 & 27 \\
    $330.952\!-\!0.182^{a}$ & Y  & 2 & Y & 5 & 130 \\
    $331.425\!+\!0.264^{a}$ & N  & 2 & N & 5 & $<$3.8 \\
    $331.120\!-\!0.118^{a}$ & N  & 2 &   & 5 & $<$3.1 \\
    $331.132\!-\!0.244$     & Y  & 2 & Y & 5 & 21 \\
    $331.278\!-\!0.188$     & Y  & 2 & Y & 5 & 95 \\
    $331.342\!-\!0.346$     & Y  & 2 & N & 5 & 60 \\
    $331.442\!-\!0.187^{a}$ & Y  & 2 & Y & 5 & 8.7 \\
    $331.542\!-\!0.066$     & Y  & 2 & N & 5 & 320 \\
    $331.556\!-\!0.121^{a}$ & Y  & 2 & Y & 5 & 106 \\
    $332.094\!-\!0.421$     & N  & 2 & Y & 5 & 480 \\
    $332.295\!-\!0.094$     & N  & 2 & Y & 5 & 71 \\
    $332.351\!-\!0.436$     & N  & 2 & N & 5 & $<$2.8 \\
    $332.560\!-\!0.148$     & N  & 2 & N & 5 & $<$11 \\
    $332.604\!-\!0.167$     & N  & 2 & N & 5 & $<$3.0 \\
    $332.942\!-\!0.686^{a}$ & N  & 2 & Y & 5 & $<$8.6 \\
    $332.963\!-\!0.679^{a}$ & N  & 2 & N & 5 & 31 \\
    $333.029\!-\!0.015$     & N  & 2 & N & 5 & $<$4.8 \\
    $333.029\!-\!0.063^{a}$ & N  & 2 &   & 5 & 32 \\
    $333.068\!-\!0.447$     & N  & 2 & N & 5 & 120 \\
    $333.121\!-\!0.434$     & N  & 2 & Y & 5 & 420 \\
    $333.128\!-\!0.440$     & N  & 2 & Y & 5 & 420 \\
    $333.130\!-\!0.560^{a}$ & N  & 2 & Y & 5 & $<$7.1 \\
    $333.163\!-\!0.101^{a}$ & N  & 2 & N & 5 & 58 \\
    $333.184\!-\!0.091$     & N  & 2 & N & 5 & $<$11 \\
    $333.234\!-\!0.062$     & Y  & 2 & Y & 5 & $<$6.0 \\
    $333.315\!+\!0.105$     & Y  & 2 & N & 5 & 56 \\
    $333.466\!-\!0.164$     & Y  & 2 & Y & 5 & 29 \\
    $333.562\!-\!0.025$     & N  & 2 & N & 5 & $<$3.9 \\ \hline
  \end{tabular}
  \label{tab:assoc}
\end{table}

\begin{table}
  \contcaption{}
  \begin{tabular}{lccccc} \hline
    {\bf Source} & \multicolumn{2}{c}{\bf 1665-MHz} & 
    \multicolumn{2}{c}{\bf 22-GHz} &
    \multicolumn{1}{c}{\bf MSX} \\
    {\bf Name}   & \multicolumn{2}{c}{\bf OH maser} & 
    \multicolumn{2}{c}{\bf \water\ maser} & 
    \multicolumn{1}{c}{\bf 21-\micron\ flux} \\
                 & {\bf Assoc} & {\bf Ref} & {\bf Assoc} & 
    {\bf Ref} & {\bf ($\mu$Wm$^{-2}$sr$^{-1}$)} \\ \hline
    $333.646\!+\!0.058^{a}$ & N  & 2 &   & 5 & $<$3.3 \\
    $333.683\!-\!0.437$     & N  & 2 & N & 5 & 1.3 \\
    $333.931\!-\!0.135$     & N  & 2 & N & 5 & 10.0 \\
    $334.635\!-\!0.015$     & N  & 2 & N & 5 & $<$1.9 \\
    $334.935\!-\!0.098^{a}$ & N  & 2 &   & 5 & $<$1.4 \\
    $335.060\!-\!0.427^{a}$ & Y  & 2 & Y & 5 & 17 \\
    $25.386\!+\!0.005$      & N  & 3 & N & 5 & $<$4.4 \\
    $25.411\!+\!0.105$      & N  & 3 & N & 5 & 28 \\
    $25.53+\!0\!.38 $       & N  & 3 & N & 5 & $<$1.7 \\
    $25.710\!+\!0.044$      & N  & 3 & N & 5 & 21 \\
    $25.826\!-\!0.178$      & N  & 3 & Y & 5 & $<$2.8 \\
    $26.528\!-\!0.266$      & N  & 3 & N & 5 & $<$7.4 \\
    $26.602\!-\!0.220$      & N  & 3 & N & 5 & $<$3.7 \\
    $27.223\!+\!0.137$      & N  & 3 & N & 5 & $<$7.4 \\
    $27.286\!+\!0.151^{a}$  & N  & 3 & N & 5 & 9.3 \\
    $27.369\!-\!0.164$      & Y  & 3 & Y & 5 & 11 \\
    $28.151\!-\!0.002$      & N  & 3 & N & 5 & $<$5.2 \\
    $28.201\!-\!0.049$      & Y  & 3 & N & 5 & 230 \\
    $28.303\!-\!0.389$      & N  & 3 & N & 5 & 120 \\
    $28.829\!+\!0.488$      & N  & 3 & Y & 5 & $<$1.8 \\
    $28.863\!-\!0.237$      & Y  & 3 & N & 5 & $<$3.6 \\
    $28.810\!+\!0.360$      & N  & 3 & Y & 5 & $<$1.7 \\
    $29.313\!-\!0.165$      & N  & 3 & N & 5 & $<$1.9 \\
    $29.867\!-\!0.042$      & Y  & 4 & N & 5 & 13 \\
    $29.865\!-\!0.007$      & N  & 3 & N & 5 & $<$2.5 \\
    $29.895\!-\!0.047$      & N  & 3 & N & 5 & $<$11 \\
    $29.907\!-\!0.040$      & N  & 3 & N & 5 & 14 \\
    $29.918\!-\!0.035$      & Y  & 4 & N & 5 & 18 \\
    $29.923\!+\!0.059$      & N  & 3 & N & 5 & $<$0.35 \\
    $29.974\!-\!0.029$      & N  & 3 & Y & 5 & 17 \\
    $30.009\!-\!0.017$      & N  & 3 & N & 5 & $<$4.7 \\
   \end{tabular}
\end{table}

\subsection{Infrared characteristics of class~I and II methanol masers}
\label{sec:infrared}

A number of searches for class~II methanol masers have been targeted
towards {\em IRAS} sources with far-infrared colours within certain
ranges \citep[e.g.][]{SVGM93,WHRB97,SVK+99,SHK00}.  The current
class~I maser observations were targeted towards class~II methanol
maser positions, the vast majority of which are known to sub-arcsecond
accuracy.  So it is possible to reliably determine whether infrared
sources from the {\em IRAS}, {\em MSX} and 2MASS point sources
catalogues are associated with the maser positions.  If the sources
with class~I methanol masers represent a different evolutionary phase
from those without then this may lead to an observable difference in
the properties of associated infrared sources.  In this section we
examine the characteristics of {\em IRAS}, {\em MSX} and 2MASS sources
associated with the masers to see if there is any difference between
those class~II methanol masers with and without associated class~I
masers.  Some readers may prefer to skip to the concluding paragraph
of this section where the results are summerised, rather read all the
details of the comparisons undertaken that are given below.

Considering only the statistically complete sample of sixty-eight
class~II methanol masers, 30 have an associated {\em IRAS} point
source \citep{IRAS} within 30 arcsec, 45 have an associated {\em MSX}
point source \citep{E+03} within 30 arcsec and 45 have an associated
2MASS source within 5 arcsec.  The names and the distance between the
infrared point source and maser positions are summarised in
table~\ref{tab:infrared}.

There are 11 {\em IRAS} sources which exhibit both class~I and II
methanol maser emission (version 2.1 of the {\em IRAS} PSC).  This is
consistent with the number expected by chance considering the relative
proportions of class~II methanol maser sources with associated class~I
masers and {\em IRAS} sources.  Examining a plot of the
$F_{25}/F_{12}$ versus $F_{60}/F_{25}$ colours (where $F_{12}, F_{25}$
and $F_{60}$ are the {\em IRAS} 12-, 25- and 60-\micron\ flux densities
respectively) there is no apparent difference between those {\em IRAS}
sources with and without an associated class~I methanol maser.  So
there doesn't appear to be any means of using {\em IRAS} colours to
select high-mass star forming regions that are more likely to have
associated class~I methanol masers, beyond the well known
ultra-compact \ionhy\ region criteria developed by \citet{WC89}.  This
implies that any evolutionary difference between class~II maser
sources with and without associated class~I masers cannot be
distinguished from {\em IRAS} data.

\begin{table*}
  \caption{Associations of the class~II methanol masers with {\em IRAS},
    {\em MSX} and 2MASS point sources.  Notes : $^{a}$ sources 
    not within the statistically complete sample of 6.6-GHz methanol masers.}
  \begin{tabular}{llrlrlr} \hline
    {\bf Source} & \multicolumn{2}{c}{\bf IRAS} & \multicolumn{2}{c}{\bf MSX} &
    \multicolumn{2}{c}{\bf 2MASS} \\
    {\bf Name}   & {\bf Name} & {\bf Distance} &{\bf Name} & {\bf Distance} &
    {\bf Name} & {\bf Distance} \\
                 &            & {\bf (arcsec)} &            & {\bf (arcsec)} & 
               & {\bf (arcsec)} \\ \hline 
    $285.32\!-\!0.03$       & $10303\!-\!5746$ & 21.4  & 
    G$285.3472\!+\!00.0013$ &  17.9 & $10321294\!-\!5802306$  &  4.8 \\
    $291.28\!-\!0.71$       & $11097\!-\!6102$ & 19.4  & 
    G$291.2731\!-\!00.7101$ &  16.0 & $11115460\!-\!6118279$  &  2.7 \\
    $293.84\!-\!0.78$       & $11298\!-\!6155$ &  1.3  & 
    G$293.8282\!-\!00.7445$ &   1.6 & $11320631\!-\!6212206$  &  2.4 \\
    $293.95\!-\!0.91$       & $11304\!-\!6206$ &  1.1  & 
    G$293.9512\!-\!00.8941$ &   6.4 & $11324316\!-\!6223048$  &  2.2 \\
    $326.475\!+\!0.703$     & $15394\!-\!5358$ & 20.8  & 
     &       & $15431671\!-\!5407089$  &  3.8 \\
    $326.641\!+\!0.613^{a}$ &                  &       & 
     &       & $15443293\!-\!5405293$  &  0.8 \\
    $326.662\!+\!0.521$     & $15412\!-\!5359$ &  6.9  & 
    G$326.6618\!+\!00.5207$ &   0.8 & $15450281\!-\!5409030$  &  0.4 \\
    $326.859\!-\!0.677^{a}$ &                  &       & 
     &       &                         &      \\
    $327.120\!+\!0.511$     & $15437\!-\!5343$ &  3.6  & 
    G$327.1192\!+\!00.5103$ &   1.9 & $15473282\!-\!5352398$  &  1.2 \\
    $327.392\!+\!0.199$     & $15464\!-\!5348$ &  4.0  & 
    G$327.3941\!+\!00.1970$ &  12.2 & $15501855\!-\!5357087$  &  2.4 \\
    $327.401\!+\!0.445$     &                  &       & 
    G$327.4014\!+\!00.4454$ &   3.4 &                         &      \\
    $327.402\!+\!0.444$     &                  &       & 
    G$327.4014\!+\!00.4454$ &   4.5 & $15491950\!-\!5345096$  &  4.5 \\
    $327.590\!-\!0.094$     &                  &       & 
     &       & $15523652\!-\!5403157$  &  4.1 \\
    $327.618\!-\!0.111$     &                  &       & 
    G$327.6184\!-\!00.1109$ &   0.4 & $15525032\!-\!5402596$  &  1.3 \\
    $327.945\!-\!0.115$     & $15507\!-\!5341$ &  2.8  & 
    G$327.9455\!-\!00.1149$ &   3.5 & $15543391\!-\!5350446$  &  0.1 \\
    $328.254\!-\!0.532$     & $15541\!-\!5349$ &  2.7  & 
    G$328.2523\!-\!00.5320$ &   5.9 &                         &      \\
    $328.237\!-\!0.548^{a}$ &                  &       & 
    G$328.2396\!-\!00.5460$ &  12.3 &                         &      \\
    $328.809\!+\!0.633^{a}$ & $15520\!-\!5234$ &  4.1  & 
    G$328.8074\!+\!00.6324$ &   4.7 & $15554848\!-\!5243018$  &  4.5 \\
    $329.031\!-\!0.198$     & $15566\!-\!5304$ &  8.4  & 
     &       & $16003085\!-\!5312268$  &  4.8 \\
    $329.029\!-\!0.205$     & $15566\!-\!5304$ & 17.6  & 
     &       &                         &      \\
    $329.066\!-\!0.308$     & $15573\!-\!5307$ &  2.2  & 
    G$329.0663\!-\!00.3081$ &   2.2 & $16010999\!-\!5316029$  &  0.6 \\
    $329.183\!-\!0.314$     & $15579\!-\!5303$ &  5.2  & 
     &       &                         &      \\
    $329.339\!+\!0.148$     & $15567\!-\!5236$ &  5.9  & 
    G$329.3371\!+\!00.1469$ &   7.7 &                         &      \\
    $329.407\!-\!0.459$     & $15596\!-\!5301$ & 13.1  & 
    G$329.4055\!-\!00.4574$ &   8.1 &                         &      \\
    $329.469\!+\!0.502$     &                  &       & 
    G$329.4626\!+\!00.5037$ &  23.2 & $15594047\!-\!5223265$  &  2.6 \\
    $329.622\!+\!0.138$     &                  &       & 
     &       & $16020013\!-\!5233578$  &  2.0 \\
    $329.610\!+\!0.114$     &                  &       & 
    G$329.6098\!+\!00.1139$ &   1.2 & $16020313\!-\!5235328$  &  0.7 \\
    $330.952\!-\!0.182^{a}$ &                  &       & 
    G$330.9544\!-\!00.1817$ &   6.8 & $16095195\!-\!5154595$  &  4.2 \\
    $331.425\!+\!0.264^{a}$ &                  &       & 
     &       &                         &      \\
    $331.120\!-\!0.118^{a}$ &                  &       & 
    G$331.1195\!-\!00.1191$ &   4.5 &                         &      \\
    $331.132\!-\!0.244$     & $16071\!-\!5142$ & 26.7  & 
    G$331.1282\!-\!00.2436$ &  12.8 & $16105979\!-\!5150253$  &  2.7 \\
    $331.278\!-\!0.188$     & $16076\!-\!5134$ & 10.4  & 
    G$331.2759\!-\!00.1891$ &   8.7 &                         &      \\
    $331.342\!-\!0.346$     &                  &       & 
    G$331.3402\!-\!00.3444$ &   8.5 & $16122648\!-\!5146177$  &  0.9 \\
    $331.442\!-\!0.187^{a}$ &                  &       & 
    G$331.4442\!-\!00.1877$ &   9.0 &                         &      \\
    $331.542\!-\!0.066$     &                  &       & 
    G$331.5414\!-\!00.0675$ &   5.1 & $16120865\!-\!5125488$  &  3.6 \\
    $331.556\!-\!0.121^{a}$ & $16086\!-\!5119$ & 26.6  & 
    G$331.5582\!-\!00.1206$ &   9.5 & $16122682\!-\!5127389$  &  3.7 \\
    $332.094\!-\!0.421$     & $16124\!-\!5110$ &  3.8  & 
    G$332.0939\!-\!00.4206$ &   1.3 & $16161646\!-\!5118251$  &  0.3 \\
    $332.295\!-\!0.094$     & $16119\!-\!5048$ &  8.1  & 
    G$332.2944\!-\!00.0962$ &   9.7 & $16154514\!-\!5055511$  &  3.4 \\
    $332.351\!-\!0.436$     & $16137\!-\!5100$ &  9.6  & 
     &       & $16173117\!-\!5108181$  &  5.0 \\
    $332.560\!-\!0.148$     &                  &       & 
    G$332.5555\!-\!00.1429$ &  23.5 &                         &      \\
    $332.604\!-\!0.167$     &                  &       & 
     &       &                         &      \\
    $332.942\!-\!0.686^{a}$ & $16175\!-\!5046$ & 23.0  & 
    G$332.9419\!-\!00.6849$ &   3.3 &                         &      \\
    $332.963\!-\!0.679^{a}$ & $16175\!-\!5045$ &  2.8  & 
    G$332.9636\!-\!00.6800$ &   4.3 &                         &      \\
    $333.029\!-\!0.015$     &                  &       & 
    G$333.0274\!-\!00.0131$ &   9.3 & $16184404\!-\!5021487$  &  2.4 \\
    $333.029\!-\!0.063^{a}$ &                  &       & 
    G$333.0299\!-\!00.0645$ &   7.1 & $16185660\!-\!5023542$  &  1.2 \\
    $333.068\!-\!0.447$     &                  &       & 
    G$333.0682\!-\!00.4461$ &   3.3 &                         &      \\
    $333.121\!-\!0.434$     &                  &       & 
    G$333.1256\!-\!00.4367$ &  20.2 & $16205937\!-\!5035516$  &  3.2 \\
    $333.128\!-\!0.440$     &                  &       & 
    G$333.1256\!-\!00.4367$ &  14.6 & $16210349\!-\!5035505$  &  2.0 \\
    $333.130\!-\!0.560^{a}$ &                  &       & 
     &       & $16213571\!-\!5040527$  &  1.5 \\
    $333.163\!-\!0.101^{a}$ & $16159\!-\!5012$ & 20.3  & 
    G$333.1642\!-\!00.0994$ &   5.6 & $16194245\!-\!5019532$  &  2.0 \\
    $333.184\!-\!0.091$     &                  &       & 
     &       & $16194611\!-\!5018365$  &  5.0 \\
    $333.234\!-\!0.062$     &                  &       & 
     &       & $16195102\!-\!5015176$  &  4.1 \\
    $333.315\!+\!0.105$     & $16157\!-\!4957$ &  2.9  & 
    G$333.3151\!+\!00.1053$ &   0.8 & $16192910\!-\!5004432$  &  2.0 \\
    $333.466\!-\!0.164$     & $16175\!-\!5002$ & 12.7  & 
    G$333.4680\!-\!00.1603$ &  16.1 &                         &      \\
    $333.562\!-\!0.025$     &                  &       & 
     &       & $16210884\!-\!4959484$  &  0.5 \\
    $333.646\!+\!0.058^{a}$ &                  &       & 
    G$333.6506\!+\!00.0598$ &  20.2 & $16210963\!-\!4952455$  &  4.8 \\
    $333.683\!-\!0.437$     & $16196\!-\!5005$ & 14.0  & 
    G$333.6788\!-\!00.4344$ &  15.5 & $16232951\!-\!5012118$  &  4.1 \\
    $333.931\!-\!0.135$     & $16194\!-\!4941$ & 25.0  & 
    G$333.9305\!-\!00.1319$ &   9.4 &                         &      \\
    $334.635\!-\!0.015$     & $16220\!-\!4906$ &  7.9  & 
    G$334.6340\!-\!00.0125$ &   8.6 & $16254596\!-\!4913388$  &  2.6 \\
    $334.935\!-\!0.098^{a}$ &                  &       & 
     &       & $16272411\!-\!4904076$  &  3.9 \\
    $335.060\!-\!0.427^{a}$ & $16256\!-\!4905$ & 17.6  & 
    G$335.0611\!-\!00.4261$ &   7.6 & $16292288\!-\!4912259$  &  3.0 \\ \hline
  \end{tabular}
  \label{tab:infrared}
\end{table*}

\begin{table*}
  \contcaption{}
  \begin{tabular}{llrlrlr} \hline
    {\bf Source} & \multicolumn{2}{c}{\bf IRAS} & \multicolumn{2}{c}{\bf MSX} &
    \multicolumn{2}{c}{\bf 2MASS} \\
    {\bf Name}   & {\bf Name} & {\bf Distance} &{\bf Name} & {\bf Distance} &
    {\bf Name} & {\bf Distance} \\
                 &            & {\bf (arcsec)} &            & {\bf (arcsec)} & 
               & {\bf (arcsec)} \\ \hline 
    $25.386\!+\!0.005$      &                  &       & 
    G$025.3865\!+\!00.0036$ &   6.8 &                         &      \\
    $25.411\!+\!0.105$      & $18345\!-\!0641$ &  4.6  & 
    G$025.4118\!+\!00.1052$ &   3.0 & $18371690\!-\!0638304$  &  2.2 \\
    $25.53+\!0\!.38 $       &                  &       & 
     &       & $18363259\!-\!0624251$  &  0.9 \\
    $25.710\!+\!0.044$      &                  &       & 
    G$025.7058\!+\!00.0403$ &  19.1 &                         &      \\
    $25.826\!-\!0.178$      &                  &       & 
     &       & $18390338\!-\!0624096$  &  3.6 \\
    $26.528\!-\!0.266$      &                  &       & 
    G$026.5254\!-\!00.2667$ &   9.6 &                         &      \\
    $26.602\!-\!0.220$      &                  &       & 
     &       &                         &      \\
    $27.223\!+\!0.137$      &                  &       & 
    G$027.2220\!+\!00.1361$ &   5.3 & $18403043\!-\!0500569$  &  1.9 \\
    $27.286\!+\!0.151^{a}$  &                  &       & 
     &       & $18403448\!-\!0457137$  &  0.3 \\
    $27.369\!-\!0.164$      & $18391\!-\!0504$ & 12.5  & 
    G$027.3652\!-\!00.1638$ &  12.1 &                         &      \\
    $28.151\!-\!0.002$      &                  &       & 
    G$028.1467\!-\!00.0040$ &  15.9 &                         &      \\
    $28.201\!-\!0.049$      & $18403\!-\!0417$ &  3.9  & 
    G$028.2007\!-\!00.0494$ &   0.7 & $18425811\!-\!0413573$  &  1.4 \\
    $28.303\!-\!0.389$      &                  &       & 
    G$028.3046\!-\!00.3871$ &  10.0 & $18442217\!-\!0417463$  &  3.0 \\
    $28.829\!+\!0.488$      &                  &       & 
     &       & $18421219\!-\!0325376$  &  4.1 \\
    $28.863\!-\!0.237$      &                  &       & 
     &       &                         &      \\
    $28.810\!+\!0.360$      &                  &       & 
     &       &                         &      \\
    $29.313\!-\!0.165$      & $18427\!-\!0320$ & 29.5  & 
     &       & $18452474\!-\!0317411$  &  4.8 \\
    $29.867\!-\!0.042$      &                  &       & 
    G$029.8620\!-\!00.0444$ &  20.7 & $18455964\!-\!0244453$  &  2.6 \\
    $29.865\!-\!0.007$      &                  &       & 
     &       &                         &      \\
    $29.895\!-\!0.047$      &                  &       & 
     &       & $18460337\!-\!0243223$  &  4.9 \\
    $29.907\!-\!0.040$      &                  &       & 
     &       & $18460346\!-\!0242394$  &  3.6 \\
    $29.918\!-\!0.035$      &                  &       & 
    G$029.9183\!-\!00.0404$ &  19.8 &                         &      \\
    $29.923\!+\!0.059$      &                  &       & 
     &       & $18454403\!-\!0239052$  &  2.7 \\
    $29.974\!-\!0.029$      &                  &       & 
    G$029.9738\!-\!00.0364$ &  28.2 & $18460869\!-\!0238420$  &  2.9 \\
    $30.009\!-\!0.017$      &                  &       & 
     &       &                         &      \\ \hline
  \end{tabular}
\end{table*}

\begin{figure}
  \psfig{file=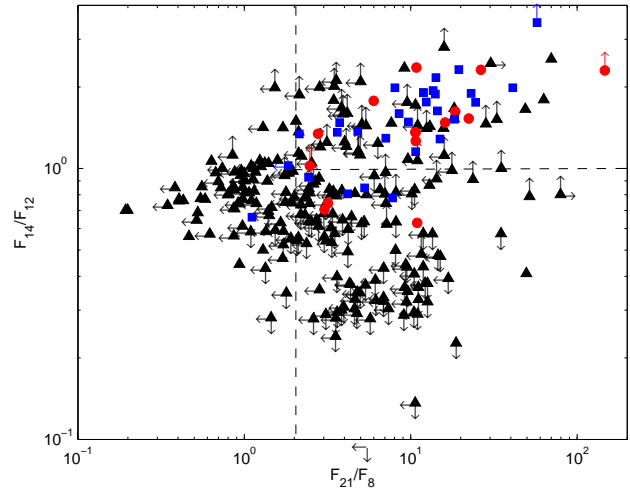,height=0.40\textwidth}
  \caption{{\em MSX} colours $F_{14}/F_{12}$ versus $F_{21}/F_{8}$ for
    class~II methanol masers with (circles) and without (squares)
    associated class~I masers.  The triangles are {\em MSX} sources
    within 30~arcsec of $l=326^{\circ}$, $b=0^{\circ}$.  Sources with
    an upper limit in one of the {\em MSX} bands making up the ratio
    are marked with an arrow pointing in the appropriate direction.
    Sources with an upper limit in both {\em MSX} bands for either of
    the colours have been excluded from the plot.  Sources within
    region within the dashed lines at the top right of the plot meet
    the MYSO criteria of \citet{LHOR02}.}
  \label{fig:msx}
\end{figure}

The {\em IRAS} observations suffered well documented problems with
confusion and saturation close to the Galactic Plane and this is
thought to be the reason why many class~II methanol maser sites have
no associated {\em IRAS} source.  Since if all class~II methanol
masers are associated with high-mass star formation (as argued the
introduction) then they should show far-infrared emission, even those
that are very young and enshrouded in cold dust.  The {\em MSX} and
2MASS observations do not have the same problems as {\em IRAS}.
However, they were made at shorter, mid- and near-infrared
wavelengths.  A total of 45 of the class~II methanol maser sources in
the statistically complete sample have an associated {\em MSX} source
within 30 arcsec, and of these 14 also have an associated class~I
methanol maser. Version 2.3 of the {\em MSX} point source catalogue
was used, which contains more than 25 percent more sources, and has
greater photometric accuracy than earlier versions of the catalogue
\citep{E+03}.  If the sources with class~I masers represent an earlier
evolutionary phase then we would expect them to be more deeply
embedded, and likely to show a lower rate of detection in the
mid-infrared {\em MSX} observations.  However, the proportion of
class~I methanol maser sources with an associated {\em MSX} source is
within the range expected given the rate of association with the
parent sample.  \citet{LHOR02} investigated the {\em MSX} colours of a
sample of massive young stellar objects (MYSO) and found that they
show $F_8 < F_{12} < F_{21}$ (where $F_{8}$, $F_{12}$ and $F_{21}$ are
the {\em MSX} 8-, 12- and 21-\micron\ flux densities respectively) and
$F_{21}/F_8 > 2$.  Figure~\ref{fig:msx} shows an MSX colour-colour
plot of $F_{14}/F_{12}$ versus $F_{21}/F_{8}$ for the class~II
methanol maser sources with and without associated class~I methanol
masers, and included for comparison are all {\em MSX} sources within
30~arcsec of $l=326^{\circ}$, $b = 0^{\circ}$.  The majority of the
maser sources meet the \citet{LHOR02} criteria for MYSO, that is they
lie within the top right section of Fig~\ref{fig:msx}.  This is not
particularly surprising as 6.6-GHz methanol masers without associated
radio continuum emission from \citet{WBHR98} were part of the sample
of sources used to define the criteria.  However, that a statistically
complete sample of class~II methanol masers shows the same
characteristics in an {\em MSX} colour-colour plot as the {\em
  IRAS}-selected sample of \citeauthor{WBHR98} adds further weight to
the widely accepted argument that the new class~II methanol masers
discovered in untargeted searches are in fact associated with
high-mass star formation.  The further towards the top-right of
Fig~\ref{fig:msx} a source lies, the cooler the implied dust
temperature, so we would expect younger, more deeply embedded sources
to lie in this region.  The class~II methanol masers with and without
associated class~I masers show the same distribution, again suggesting
that there is no significant evolutionary difference between the two
groups.

The quoted astrometric accuracy of the {\em MSX} point source
catalogue is better than 2 arcsec \citep{E+03}.  However, only 8 of
the 45 {\em MSX} sources are within 2 arcsec of the class~II maser
position.  In total 14 of the class~II maser positions are within
5~arcsec and 29 are within 10~arcsec of an {\em MSX} point source.
This suggests that in general the masers are offset from the {\em MSX}
point sources, and probably associated with other objects within the
larger star formation complex.  It suggests that in many cases the
colours of the {\em MSX} sources in Fig.~\ref{fig:msx} are not those
of the source exciting the methanol masers and so will confuse any
attempt to find differences between the class~II masers with and
without associated class~I masers.  However, it still may be possible
to find such differences directly from the {\em MSX} image data.  I
have examined the 21-\micron\ {\em MSX} (E-band) images in the
vicinity of each of the maser sources and the results are summarise in
Table~\ref{tab:assoc}.  If the class~I masers are associated with a
generally earlier evolutionary phase, then we would expect a lower
percentage to be projected against 21-\micron\ emission.  Considering
the sixty-eight class~II masers in the complete sample, thirty-five
are projected against 21-\micron\ emission, 15 of 26 with class~I
masers and 21 of 42 without.  So approximately 50 percent of class~II
methanol maser sources are projected against 21-\micron\ {\em MSX}
emission, but there is no correlation with the presence or absence of
associated class~I masers.  Of those maser sources that are not
projected against 21-\micron\ emission, some are near a source, but
the majority are not. In contrast, for an {\em IRAS}-selected sample
of nearly 50 UC\ionhy\ regions, only one did not have coincident
21-\micron\ {\em MSX} emission \citep{CC03}.

Near-infrared observations of methanol maser sources are generally of
limited use as most sources are thought to be optically thick at these
wavelengths.  However, for completeness the association of the
methanol masers with sources in the 2MASS point source catalogue has
been examined.  Considering the statistically complete sample of
sixty-eight sources, forty-five have a 2MASS point source within
5~arcsec, dropping to 13 within 2~arcsec.  The proportion of these
sources which also have an associated class~I methanol maser matches
the proportions for the sample as a whole.  As for the {\em MSX} sources,
it is likely that in many cases the 2MASS point sources are not
directly associated with the masers, but rather a nearby source within
the same region.

The association of the class~II methanol masers in the statistically
complete sample with {\em IRAS}, {\em MSX} and 2MASS sources has been
investigated.  There is no measurable difference, either in terms of
rates of association, or the infrared colours between those class~II
masers with an associated class~I maser and those without.  Combining
this with the similar finding for the association of other maser
species, it suggests that class~I methanol masers, like class~II, are
associated with star formation regions for a moderately long
evolutionary period.  For example $329.031\!-\!0.198$ is optically
thick at a wavelength of 21-\micron, suggesting it is deeply embedded
and at an early evolutionary phase.  In contrast $328.809\!+0\!.633$
has a well developed \ionhy\ region with some extended emission
\citep*{ESK05} and shows emission in a number of the rarer excited OH
and class~II maser transitions \citep[][and references
therein]{ECJ+04}, which are believed to be associated with more
evolved regions.  The lack of any distinguishing characteristics
between the class~II masers with and without associated class~I masers
is consistent with the general assumption that the two classes of
methanol maser are not directly associated.  High-mass stars form in
clusters and so it is likely that at any one time there are stars at a
variety of evolutionary phases within the one region.  If in general
the star exciting the class~II methanol masers is not the source of
the outflow producing the class~I masers then we would not expect any
clear evolutionary relationship to be manifest.  However, there are
many other possible complicating factors, for example there is no
reason to expect both classes of methanol maser to be associated with
exactly the same stellar mass range.  This complexity further
highlights the general need for high-resolution observations to
disentangle a high-mass star formation complex.

\section{Conclusions}

Class~I methanol masers are associated with approximately half of all
class~II methanol maser sources.  In contrast to previous suggestions,
there is no anti-correlation between the velocity range of the two
maser classes, nor their peak flux densities.  The velocity ranges
overlap in the majority of sources and there is some evidence that in
those sources where there is an overlap the peak flux density of the
class~I masers is stronger.  The peak flux density of the 6.6- and
95.1-GHz transitions in most sources is of the same order of
magnitude.  This suggests that the peak maser flux density in both
transitions may be heavily influenced by a common factor, such as the
general methanol abundance within the larger star formation region.

Interferometric observations of the class~I masers are required to
allow a more detailed examination of the relationship between the two
methanol maser classes and their role in the larger high-mass star
formation picture.  In particular to determine if the overlap in the
velocity ranges seen in many sources is associated with coincident or
near-coincident emission from the two transitions, or is merely an
artefact of turbulent velocity fields.

Investigation of other maser species and infrared sources associated
with the methanol masers did not find any statistically significant
correlations that can be used to target future class~I maser searches.
The absence of such correlations is consistent with the hypothesis
that the objects responsible for producing class~I methanol masers are
in general not those that produce main-line OH, water or class~II
methanol masers, although there are other possible explanations.

\section*{Acknowledgements}

Thanks to Robina Otrupcek for her assistance with observations.
Thanks to Jim Caswell for valuable comments and discussions.
Financial support for this work was provided by the Australian
Research Council.  This research has made use of NASA's Astrophysics
Data System Abstract Service and data products from the {\em Midcourse
  Space Experiment}.  Processing of the data was funded by the
Ballistic Missile Defence Organization with additional support from
the NASA Office of Space Science.  The research has made used of the
NASA/IPAC Infrared Science Archive, which is operated by the Jet
Propulsion Laboratory, California Institute of Technology, under
contract with the National Aeronautics and Space Administration.

\end{document}